\documentclass[twocolumn,showpacs,preprintnumbers,nofootinbib]{revtex4}

\usepackage{latexsym,amssymb,amsmath,bbm,graphicx,here,epsfig}
\usepackage{feynmf}
\usepackage{psfrag} 

\allowdisplaybreaks[1] 

\newcommand{\Tr}{\mathrm{Tr}}\newcommand{\STr}{\mathrm{STr}}

\newcommand{\Bs}{\boldsymbol}

\renewcommand{\bar}{\overline}

\begin{document}

\preprint{HD-THEP-03-35}
\title{Temperature dependence of antiferromagnetic order in the \\ Hubbard model}

\author{Tobias Baier}
\email{baier@thphys.uni-heidelberg.de}
\author{Eike Bick}
\email{bick@thphys.uni-heidelberg.de}
\altaffiliation[Now at ]{d-fine GmbH}
\author{Christof Wetterich}%
\email{C.Wetterich@thphys.uni-heidelberg.de}
\affiliation{%
Institut  f\"ur Theoretische Physik\\
  Universit\"at Heidelberg\\
  Philosophenweg 16, D-69120 Heidelberg
}%

\begin{abstract}
We suggest an approximative solution of the two dimensional Hubbard model close
to half filling. It is based on partial bosonization, supplemented by an investigation
of the functional renormalization group flow. The inclusion of the fermionic and bosonic
fluctuations leads in lowest order to agreement with Hartree-Fock result or
Schwinger-Dyson equation and cures the mean field ambiguity. We compute the temperature
dependence of the antiferromagnetic order parameter and the gap below the critical
temperature. We argue that the
Mermin--Wagner theorem is not practically applicable for the spontaneous
breaking of the continuous spin symmetry in the antiferromagnetic state of the
Hubbard model. The long distance behavior close to and below the critical temperature is
governed by the renormalisation flow for the effective interactions of composite Goldstone
bosons and deviates strongly from the Hartree-Fock result.
\end{abstract}

\maketitle




\section{Introduction}

The rich behaviour of strongly coupled electron systems has attracted the attention of physicists for many decades. One of the most popular models, which is believed to contain the most relevant features to explain as different phenomena as metal-insulator transitions, ferromagnetism, antiferromagnetism and, more recently, the behaviour of high temperature superconductors, is the Hubbard model \cite{Hub63}. It combines formal simplicity with a rich and complex spectrum of its implications that --- even after nearly 40 years --- are not fully understood. Analytical results are scarcely available. Only in one dimension or in higher dimensions for very special choices of the parameters exact analytical solutions of the model are known \cite{TasXX}. Besides many other attempts the last years witnessed strong efforts to investigate the properties of the model by means of renormalisation group equations \cite{Zan97,Hal00,Hon01,Gro01}. Currently, although far from being able to fully and uniquely describe the properties of the Hubbard model, renormalisation group approaches are widely believed to be a promising method to further investigate its interesting phase diagram.

Most of the renormalisation group approaches have studied the evolution of the four fermion interaction
in dependence on a renormalisation scale $k$.  Phase transitions and spontaneous symmetry breaking are regarded to be encoded in the momentum structure of this vertex.
More precisely, a phase transition is inferred from a divergence of the running coupling
at some characteristic scale $k_{SSB}$. The particular momentum structure of the
divergent part of the vertex indicates the relevant order parameter for the phase
transition.
Even though the detection of a transition by this method looks rather convincing, the appearance of a divergence at $k_\text{SSB}$ forbids a continuation of the flow for $k<k_\text{SSB}$. In consequence, it is not possible to penetrate the low temperature phase with non vanishing order parameter by a straightforward study of the four fermion
interaction.

For the two dimensional model this shortcoming is particularly annoying since an exact theorem by Mermin and Wagner \cite{Mer66} forbids the spontaneous breaking of a continuous symmetry at non zero temperature. However, the widely believed antiferromagnetism in the low temperature phase of the two dimensional Hubbard model corresponds precisely to the spontaneous breaking of the global spin rotation symmetry. Antiferromagnetic order at nonzero temperature therefore seems to contradict the exact theorem. The present paper will resolve this discrepancy and presents a computation of the antiferromagnetic order parameter for temperatures below the critical temperature.
A summary of our results can be found in \cite{BBWL}. 

Our method is based on partial bosonisation, whereby the fermion bilinear $\sim \bar\psi\psi$ is treated as a (composite) bosonic field $\phi$. For the Hubbard model, partial bosonisation replaces the four fermion interaction $\sim U(\bar\psi\psi)^2$ by a Yukawa coupling to the boson $\sim \bar h(\bar\psi\psi)\phi$ and a boson mass term $\sim\bar m^2\phi^2$, whereby $U\sim \bar h^2/\bar m^2$. The renormalisation flow describes how the mass term $\bar m^2(k)$ depends on the renormalisation scale $k$. The onset of spontaneous symmetry breaking at $k=k_\text{SSB}$ is signalled by the vanishing of $\bar m^2(k_\text{SSB})$. This indeed corresponds to a divergent four fermion coupling $U$. However, in the partially bosonised formulation it is comparatively easy to include into the truncation also the bosonic interactions generated by the flow, e.g. a quartic scalar interaction $\sim\bar\lambda \phi^4$. (In the fermionic language this corresponds to an eight--fermion interaction which is quite difficult to compute.) For positive $\bar\lambda$ a vanishing mass term $\bar m^2=0$, or even a negative mass term $\bar m^2<0$, still lead to a well behaved effective scalar potential (or free energy density)
$U_\text{eff}=\bar m^2\phi^2/2+\bar\lambda\phi^4/8$.
The minimum of the potential shifts for $\bar m^2<0$ simply to a nonvanishing value $\phi^2_0=-2\bar m^2/\bar\lambda$. The value of $\phi_0$ can be associated with the order parameter responsible for spontaneous symmetry breaking.
For temperatures $T$ below the critical temperature $T_c$ one finds that $\phi(k)$ starts
being different from zero precisely for $k=k_{SSB}>0$.
It becomes obvious that the partially bosonised version of the renormalisation group equation has no problem to describe the flow also for $k<k_\text{SSB}$ and therefore to penetrate the low temperature phase with spontaneous symmetry breaking.

This allows us to study explicitely the issue of the Mermin--Wagner theorem. Our findings are indeed consistent with the formal correctness of this theorem. In an infinitely extended space all fluctuations are included only if the effective infrared cutoff $k$ is sent to zero. Indeed, we observe that $\phi_0(k\to 0) = 0$ such that in the formal limit $k\to 0$ the order parameter vanishes and the Mermin--Wagner theorem is obeyed. However, in the low temperature phase below the critical temperature $T_c$ the ``running'' of the $k$ dependent order parameter towards zero is extremely slow -- it is only logarithmic. As a consequence we find antiferromagnetic order at all realistic macroscopic scales. We conclude that the Mermin--Wagner theorem is not practically applicable here, despite its formal correctness.

\begin{figure}[htb]
\centering
\psfrag{T}{$T/t$}
\psfrag{kappa}{$\kappa$}
\includegraphics[scale=0.65]{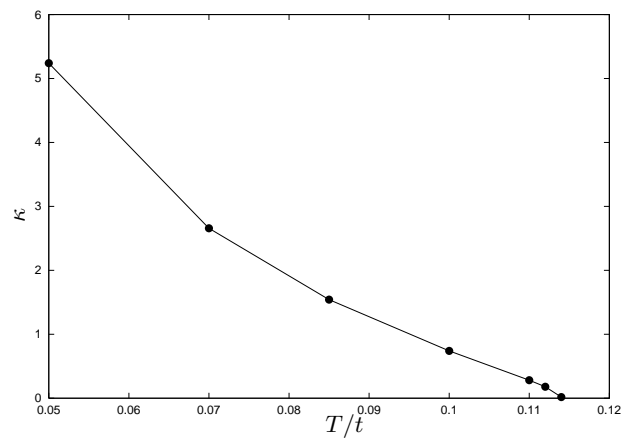}

\caption{Temperature dependence of the antiferromagnetic order parameter. We plot
the renormalized expectation value $\kappa$ of the antiferromagnetic condensate in the
low temperature phase for $U/t=3$. The scale $k/t=10^{-9}$ corresponds
to a probe of macroscopic size of roughly 1 cm. The gap for the fermionic excitations is
related to $\kappa$ by $\Delta_a/t=\sqrt{2h^2_a\kappa}$.}
\label{fig:ssb}
\end{figure}

We demonstrate this explicitely by computing the antiferromagnetic
order parameter $\kappa$ and gap $\Delta_a$ at a macroscopic scale $k^{-1}=1\text{cm}$.
The gap is directly connected to a nonvanishing antiferromagnetic order
parameter $\alpha_0\sim\phi_0^2$ by
$\Delta_a=(2\bar{h}^2_a\alpha_0)^{1/2}=(2h^2_a\kappa)^{1/2}t$.
Here $\kappa$ is the dimensionless
renormalized order papameter \footnote{Of course, in a finite volume $V\sim k^{-3}$ the usual finite volume effects also formally lead to a vanishing order parameter. This is common to all other systems with spontaneous symmetry breaking. It is a well understood effect of no relevance to our issue.}
and $h_a$ the dimensionless renormalized Yukawa coupling
such that the gap is proportional to the hopping parameter $t$ of the Hubbard model.
Fig.\ref{fig:ssb} shows the temperature dependence of the antiferromagnetic order parameter $\kappa$ at half filling (vanishing chemical potential $\mu=0$). This figure may be regarded as a central result of our paper. We clearly observe the onset of spontaneous symmetry breaking for $T<T_c$. We find that the phase transition is continuous. For this plot we have concentrated only on the simplest version of the Hubbard model which includes only next neighbour hopping with coupling strength $t$. For a realistic value $t=0.3\text{eV}$ the critical temperature corresponds to $T_c=400\text{K}$.

An investigation of the antiferromagnetic phase within the partially bosonised formulation of the Hubbard model has already been performed in \cite{Bai00} where we used mean field theory (MFT). Unfortunately, we have found that in the MFT approximation the results for $T_c$ or the antiferromagnetic order parameter depend strongly on unphysical parameters which describe the precise choice of the mean field. Since this ambiguity is closely related to the possibility of the Fierz reordering of the local four fermion interaction it is dubbed ``Fierz--ambiguity''. In the present paper we show that the inclusion of the bosonic fluctuations in the renormalisation flow (beyond MFT!) greatly reduces this Fierz--ambiguity.

In our present truncation the resolution of the momentum dependence of the effective four fermion vertex depends on the choice of composite bosons included explicitely.
More precisely, the effective vertex arises from the exchange of the bosons and therefore
inherits its momentum structure from the momentum dependence of the bosonic
propagators.
In this paper we concentrate on the antiferromagnetic order and small doping. We only take into account bosons describing spin densities and charge densities. We emphasize, however, that our formalism is quite general. It is also well adapted to include the interesting superconducting channel.

This paper is organised as follows: In sect. \ref{sec:Hubmod} we review the functional integral description of the Hubbard model and set up our notation. We proceed in sects. \ref{sec:schwinger} and \ref{sec:antifer_gap} to derive and solve the Schwinger-Dyson gap equation. In sect. \ref{sec:partbos} we introduce our partial bosonisation. We carry out a mean field calculation in sect. \ref{sec:MF} that can be directly compared to the Schwinger-Dyson result. The ``Fierz--ambiguity'' arising in this approach is described in sect. \ref{sec:fierz}. In sect. \ref{sec:bosprop} we turn to the momentum dependence of the effective bosonic propagators which is first computed in the mean field approximation. The one loop corrections to the bosonic propagators show that indeed antiferromagnetism is the dominant way of breaking the $SU(2)$-spin-symmetry of the model. The last part contains our renormalisation group approach to the Hubbard model in its partially bosonised version and particularly a discussion of the behaviour of the flow in the phase of broken symmetry. We review the exact functional renormalisation group equation for the effective average action in sect. \ref{sec:ERGE} and the rebosonisation of fermionic interactions in sect. \ref{sec:rebos}. Sect. \ref{sec:trunc_antiferr} sets up our truncation and choice of cutoff functions for the Hubbard model. In sect. \ref{sec:halffill} we derive the non--perturbative flow equations at half filling ($\mu=0$). Their solution in sect. \ref{sec:spont_symm_break} leads to an understanding of the appearance of antiferromagnetic order despite the Mermin--Wagner theorem and is the basis for our main result shown in fig. \ref{fig:ssb}.
The critical behavior is discussed in more detail in section \ref{sec:criticalbehavior}.
In sect. \ref{sec:flownonzero} we turn to nonzero doping and discuss the phase diagram of the Hubbard model for small values of the chemical potential. Section \ref{sec:remove_ambig} finally discusses how the Fierz ambiguity is substantially reduced if the effects of bosonic fluctuations are included beyond mean field theory. Our conclusions are summarised in sect. \ref{sec:conc}. In the appendices we give explicit formulae for the flow of the couplings in different truncations and show how nonlocal operators corresponding to a d--wave superconducting state may be incorporated into the formalism.

\section{The Hubbard model}
\label{sec:Hubmod}

The partition function of the Hubbard model is given by
\begin{eqnarray} \label{eq:Zorig}
Z[\eta]&=&\int_{\hat{\psi}(\beta)=-\hat{\psi}(0),\hat{\psi}^*(\beta)=-\hat{\psi}^*(0)}
{\cal D}(\hat{\psi}^*(\tau),\hat{\psi}(\tau))  \nonumber\\
&&\exp\bigg(-\int^\beta_0d\tau\bigg(\sum_{\bf{x}}\hat{\psi}^\dagger_{\bf{x}}(\tau)
\bigg(\frac{\partial}{\partial\tau}-\mu\bigg)\hat{\psi}_{\bf{x}}(\tau)\nonumber\\
&&\qquad +\sum_{\bf{xy}}\hat\psi^\dagger_x(\tau)
{\cal T}_{\bf{xy}}\hat{\psi}_{\bf{y}}(\tau)\nonumber\\
&&\qquad+\frac{1}{2}U\sum_{\bf{x}}\bigg(\hat{\psi}^\dagger_{\bf{x}}(\tau)\hat{\psi}_{\bf{x}}
(\tau)\bigg)^2\\
&&\qquad-\sum_{\bf{x}}(\eta^\dagger_{\bf{x}}(\tau)\hat{\psi}_{\bf{x}}(\tau)
+\eta^T_{\bf{x}}(\tau)\hat{\psi}^*_{\bf{x}}(\tau))\bigg)\bigg)\nonumber
\end{eqnarray}
Here
\begin{equation}
  \hat\psi_{\Bs{x}}(\tau)=\begin{pmatrix}\hat\psi_{\Bs{x}\uparrow}(\tau)\\\hat\psi_{\Bs{x}\downarrow}(\tau)\end{pmatrix},
  \quad\hat\psi_{\Bs{x}}^\dagger(\tau)=\begin{pmatrix}\hat\psi_{\Bs{x}\uparrow}^*(\tau),&\hat\psi_{\Bs{x}\downarrow}^*(\tau)\end{pmatrix}
\end{equation}
are Grassmann fields describing electrons on a quadratic lattice with lattice sites at $\Bs{x}\in\mathbbm{Z}\times\mathbbm{Z}$. (We restrict our attention to the two dimensional case.)
The Euclideon time $\tau$ parameterises a torus with circumference $\beta=T^{-1}$.
The functional integral is constrained by antiperiodic boundary conditions due to the fermionic character of the fields. The kinetic term contains the chemical potential $\mu$ as well as the so called {\em hopping matrix} $\cal T$, which describes the possibility of electron hopping between different lattice sites. For simplicity, we assume
\begin{equation}
  {\cal T}_{\Bs{x}\Bs{y}}=\left\{\begin{aligned}-t&\quad\text{, if $\Bs{x}$ and $\Bs{y}$ are nearest neighbors}\\
      0 &\quad\text{, else.}\end{aligned}\right.
\end{equation}
The quartic term reflects the local Coulomb repulsion of electrons at the same lattice site. Additionally we include fermionic sources $\eta_{\Bs{x}}^*(\tau)$, $\eta_{\Bs{x}}(\tau)$.

Calculations are conveniently carried out in momentum space. We define the Fourier transforms
\begin{equation}
  \label{eq:four}
  \begin{aligned}
    \hat\psi(X)&=\sum_Q\hat\psi(Q)e^{iQX},\quad
    \hat\psi^*(X)&=\sum_Q\hat\psi^*(Q)e^{-iQX},
  \end{aligned}
\end{equation}
where we use the short hand notations.
\begin{equation}
  \begin{gathered}
    Q=(\omega_F,\Bs{q}),\quad X=(\tau,\Bs{x}),\quad QX=\omega_F\tau+\Bs{x}\Bs{q}
  \end{gathered}
\end{equation}
in position and momentum space and define sums and delta functions accordingly
\begin{equation}
  \begin{gathered}
    \sum_X=\int_0^\beta d\tau\,\sum_{\Bs{x}},\quad\sum_Q=T\sum_{n=-\infty}^\infty\int_{-\pi}^\pi\frac{d^2q}{(2\pi)^2},\\
    \delta(Q-Q')=\beta\delta_{n,n'}(2\pi)^2\delta(\Bs{q}-\Bs{q'}),\\
    \delta(X-X')=\delta(\tau-\tau')\delta_{\Bs{x},\Bs{x'}}.
  \end{gathered}
\end{equation}
Here $\omega_F=(2n+1)\pi T$ is called {\em Matsubara frequency} and the corresponding sum {\em Matsubara sum}, with inverse temperature $\beta=T^{-1}$. All components of $Q$ or $X$ aquire the same canonical dimension if $\Bs q$ and $\Bs x$ are measured in units of the lattice distance, i.e. $\Bs x\to a\Bs x$, $\Bs q \to \Bs q/a$. Here we set $a=1$. Note that all functions in momentum space are $2\pi$-periodic in spatial momentum. In particular, the kinetic term in momentum space takes the form
\begin{equation}\label{7}
  \begin{split}
    S_{F,\text{kin}}&=\sum_Q\hat\psi^\dagger(Q)P(Q)\hat\psi(Q),\\
    P(Q)&=i\omega_F-\mu-2t(\cos q_1+\cos q_2),
  \end{split}
\end{equation}
such that the Fermi surface is located at $-2t(\cos q_1+\cos q_2)=\mu$.

Another convenient notation that we will adopt is to introduce a collective index $\alpha$ denoting spin, $X$ or $Q$, and the label for the two complex conjugated fermionic fields. Correspondingly, $a=1\dots 4$ is the collective index associated with the discrete
``internal'' indices only. Explicitly, we have
\begin{equation}
  \hat\psi_a(X)=\begin{pmatrix}\hat\psi_\uparrow\\\hat\psi_\downarrow\\\hat\psi_\uparrow^*\\\hat\psi_\downarrow^*\end{pmatrix}_a(X).
\end{equation}
In this notation, the partition function reads
\begin{equation}
  \label{eq:Zschwdys}
  Z[\eta]=\int{\cal D}\hat\psi_\alpha\exp\left(-S[\hat\psi]+\eta_\alpha\hat\psi_\alpha\right)
\end{equation}
with the action
\begin{equation}
  \label{eq:Sschwdys}
  S[\hat\psi]=-\frac{1}{2}j_{\alpha\beta}\hat\psi_\alpha\hat\psi_\beta+\frac{1}{24}\lambda_{\alpha\beta\gamma\delta}\hat\psi_\alpha\hat\psi_\beta\hat\psi_\gamma\hat\psi_\delta.
\end{equation}
In momentum space $j_{\alpha\beta}=j_{ab}(P,Q)$ is defined by the Fourier transforms \eqref{eq:four}, implying
\begin{equation}
  \label{eq:classprop}
  \begin{split}
    j_{ab}(P,Q)&\equiv J_{ab}(Q)\delta(P-Q),\\
    J(Q)&=\begin{pmatrix}0&P(-Q)\mathbbm{1}_2\\-P(Q)\mathbbm{1}_2&0\end{pmatrix}.
  \end{split}
\end{equation}
The quartic coupling $\lambda_{\alpha\beta\gamma\delta}$ is most easily given in position space. By comparing \eqref{eq:Zschwdys} with \eqref{eq:Zorig} we get
\begin{equation}
  \label{eq:quartcoup}
  \lambda_{\alpha\beta\gamma\delta}=-U\epsilon_{abcd}\,\delta(X-X_2)\delta(X-X_3)\delta(X-X_4)
\end{equation}
with $\alpha=(X,a)$, $\beta=(X_2,b)$, $\gamma=(X_3,c)$ and $\delta=(X_4,d)$.

\section{Schwinger-Dyson equation}
\label{sec:schwinger}

Starting from the partition function \eqref{eq:Zorig}, we derive the Schwinger-Dyson
equation \cite{SD} for the fermionic two point function. The lowest order approximation to the Schwinger-Dyson equation is the Hartree-Fock equation \cite{HFH,MFL,HF,MFH}. We start by defining the {\em effective action} by the Legendre transform of the logarithm of the partition function in the standard way:
\begin{equation}
  \label{eq:defEA}
  \Gamma[\psi]=-\ln Z[\eta[\psi]]+\eta_\alpha[\psi]\psi_\alpha.
\end{equation}
Here $\psi_\alpha=\langle\hat\psi_\alpha\rangle=\frac{\partial}{\partial\eta_\alpha}\ln Z[\eta]$ denotes the ``expectation value'' of the corresponding fermionic field in presence of fermionic sources. It is possible to rewrite the effective action
as an implicit functional integral
\begin{equation}\label{eq:14}
  \Gamma[\psi]=-\ln\int{\cal D}\tilde\psi\,\exp\left(-S[\psi+\tilde\psi]+\eta_\alpha\tilde\psi_\alpha\right)
\end{equation}
with $\eta_\alpha=-\partial\Gamma/\partial\psi_\alpha$. We will use this as
the starting point for the derivation of the Schwinger-Dyson equation
\cite{Wet02}. For vanishing sources one finds the derivative of eq. (\ref{eq:14})
\begin{eqnarray}\label{eq:schwdys1}
&&\frac{\partial\Gamma[\psi]}{\partial\psi_\sigma}=-j_{\sigma\alpha}\psi_\alpha+\frac{1}{6}
\lambda_{\sigma\alpha\beta\gamma}\langle\hat\psi_\alpha\hat\psi_\beta\hat\psi_\gamma\rangle\\
&&\hspace{1.05cm}=-j_{\sigma\alpha}\psi_\alpha+\frac{1}{6}\lambda_{\sigma\alpha\beta\gamma}
\Big(\psi_\alpha\psi_\beta\psi_\gamma\nonumber\\
&&+(\Gamma^{(2)})^{-1}_{\alpha\beta}\psi_\gamma
-(\Gamma^{(2)})^{-1}_{\alpha\gamma}\psi_\beta
+(\Gamma^{(2)})^{-1}_{\beta\gamma}\psi_\alpha\nonumber\\
&&-(\Gamma^{(2)})^{-1}_{\alpha\alpha'}(\Gamma^{(2)})^{-1}_{\beta\beta'}
(\Gamma^{(2)})^{-1}_{\gamma\gamma'}
\frac{\partial^3\Gamma}{\partial\psi_{\alpha'}
\partial\psi_{\beta'}\partial\psi_{\gamma'}}\Big),\nonumber
\end{eqnarray}
where
\begin{equation}
  (\Gamma^{(2)})_{\alpha\beta}=-\frac{\partial^2\Gamma}{\partial\psi_\alpha\partial\psi_\beta}
\end{equation}
is the inverse fermionic propagator. Equation \eqref{eq:schwdys1} can be read as an identity relating various fermionic $n$-point functions. Differentiating \eqref{eq:schwdys1} once more and setting $\psi=0$, we find the Schwinger-Dyson equation
\begin{eqnarray}
&&\Gamma_{\rho\sigma}^{(2)}=-j_{\rho\sigma}+\frac{1}{2}\lambda_{\rho\sigma\alpha\beta}
(\Gamma^{(2)})^{-1}_{\alpha\beta}\\
&&+\frac{1}{6}\lambda_{\sigma\alpha\beta\gamma}(\Gamma^{(2)})^{-1}_{\alpha\alpha'}
(\Gamma^{(2)})^{-1}_{\beta\beta'}(\Gamma^{(2)})^{-1}_{\gamma\gamma'}
\frac{\partial^4\Gamma}{\partial\psi_\rho\partial\psi_{\alpha'}\partial\psi_{\beta'}
\partial\psi_{\gamma'}}  \nonumber
\end{eqnarray}
relating the fermionic 2- and 4-point functions. The last term in this equation represents a two loop expression of the order $\lambda^2$ \cite{Wet02} and will  be neglected in this and the next section.The lowest order Schwinger--Dyson equation results in a closed equation
\begin{equation}
  \label{eq:schwdys}
  G_{\alpha\beta}^{-1}=-j_{\alpha\beta}+\frac{1}{2}\lambda_{\alpha\beta\gamma\delta}G_{\gamma\delta}
\end{equation}
for the two point function $G_{\alpha\beta}$,
\begin{equation}
  G_{\alpha\beta}=(\Gamma^{(2)})^{-1}_{\alpha\beta}.
\end{equation}
It can be integrated in order to obtain the two particle irreducible effective action or
bosonic effective action \cite{LWB}, \cite{Wet02} and a corresponding expression for the
free energy. 

\section{Antiferromagnetic gap equation}
\label{sec:antifer_gap}

The Hubbard model \eqref{eq:Zorig} is symmetric under global $SU(2)$-spin - and $U(1)$-charge-transformations. However, for small T and small $\mu$ (corresponding to small doping) it is well known that the Hubbard model exhibits antiferromagnetic behaviour. A nonvanishing expectation value of the operator $\hat\psi^\dagger(X)\vec\sigma\hat\psi(X)$ in macroscopically large regions can be associated with spontaneous breaking of the $SU(2)$--spin--symmetry. Even though spontaneous symmetry breaking of a continuous symmetry is forbidden in a strict sense by the Mermin--Wagner theorem \cite{Mer66} we will see in section \ref{sec:spont_symm_break} that the main physical features are correctly described by such a ``naive spontaneous symmetry breaking''. The leading order Schwinger--Dyson equation exhibits a nonzero expectation value\footnote{Note that the result of the leading order Schwinger--Dyson equation is not consistent with the Mermin--Wagner theorem. Our more extended approximations based on the renormalisation group equation will establish the missing consistency.} of the antiferromagnetic condensate\footnote{The Yukawa coupling $h_a$ is introduced solely for consistency with later notation and plays no role here, only $h_a\vec a$ being observable.}
\begin{equation}
  \label{eq:gapfun_a}
  g^{(a)}_{ab}(X) = \frac{h_a\vec a}{U}\exp(i\Pi X)
  \begin{pmatrix} 0&-\vec\sigma\\\vec\sigma^T&0 \end{pmatrix},
\end{equation}
with $\Pi=(0,\pi,\pi)$. Here the factor $\exp(i\Pi X)$ ensures the alternating sign of the spin density between neighbouring lattice sites. We therefore investigate the possible solutions of equation \eqref{eq:schwdys} for Green functions obeying
\begin{equation}
  G_{ab}(X,X) = g^{(h)}_{ab}(X) + g^{(a)}_{ab}(X)
\end{equation}
where the symmetry conserving part can be parameterised as
\begin{equation}
  \label{eq:gabfun_h}
  g^{(h)}_{ab}(X) = - \frac{h_\rho \rho}{U} \begin{pmatrix} 0&-\mathbbm{1}_2\\\mathbbm{1}_2&0 \end{pmatrix}.
\end{equation}
Since the second term on the r.h.s. of equation \eqref{eq:schwdys} is local (due to equation \eqref{eq:quartcoup}) it only involves $G_{ab}(X,X)$ and therefore induces a ``local gap''
\begin{equation}
  \begin{aligned}
    \Delta_{ab}(X) &= -\frac U2 \epsilon_{abcd}(g^{(h)}_{cd}(X)+g^{(a)}_{cd}(X))\\
    &=(\Delta_\rho)_{ab}+(\bar{\Delta}_a)_{ab}(X)
  \end{aligned}
\end{equation}
with {\em charge density shift}
\begin{equation}
  \label{eq:rhogap}
  \Delta_\rho=-h_\rho\rho\begin{pmatrix}0&-\mathbbm{1}_2\\\mathbbm{1}_2&0\end{pmatrix}
\end{equation}
and {\em antiferromagnetic spin density gap}
\begin{equation}
  \label{eq:agap}
  \begin{aligned}
    (\bar{\Delta}_a)_{ab}(X)&=(\Delta_a)_{ab}\exp(i\Pi X),\\
    \Delta_a&=-h_a\begin{pmatrix}0&-\vec\sigma^T\vec a\\\vec\sigma\vec a&0\end{pmatrix}.
  \end{aligned}
\end{equation}
For nonvanishing $\Delta_a$ the fermionic propagator indeed acquires a mass gap according to
\begin{equation}
  \label{eq:gapansatz}
  G_{ab}^{-1}(X,Y)=-j_{ab}(X,Y)+\Delta_{ab}(X)\delta(X-Y).
\end{equation}
On the other hand, we will see below that the charge density shift only shifts the effective
value of the chemical potential, $\mu_{eff}=\mu=h_\rho\rho(\mu)$ with $\rho(\mu=0)=0$.

In order to determine $h_a\vec a$ and $h_\rho\rho$ as functions of $T$ and $\mu$ we have to solve the ``gap equation'' which obtains by evaluating the inverse of equation \eqref{eq:gapansatz} at equal arguments $X=Y$, i.e.
\begin{equation}
  g(X)=G(X,X)=(-j+\Delta)^{-1}(X,X).
\end{equation}
In momentum space this becomes
\begin{equation}
  \label{eq:schwdyslhs}
  g(Q)=\sum_KG(K,Q+K)=\sum_K(-j+\Delta)^{-1}(K,Q+K)
\end{equation}
with gap function
\begin{equation}
  \Delta(P,Q)=\Delta_\rho\delta(P-Q)+\Delta_a\delta(P-Q+\Pi ).
\end{equation}
The solution of \eqref{eq:schwdyslhs} needs the inversion of the inverse propagator \eqref{eq:gapansatz} in momentum space
\begin{equation}
  \label{eq:gapansatz1}
  (-j+\Delta)(P,Q)=(-J(P)+\Delta_\rho)\delta(P-Q)+\Delta_a\delta(P-Q+\Pi ).
\end{equation}
By use of the identity
\begin{equation}
  \Delta_a^*(-J(Q)+\Delta_\rho)^*=(-J(Q)+\Delta_\rho)\Delta_a
\end{equation}
we find
\begin{equation}
  \label{eq:inverse_propagator}
  \begin{split}
    (-j+&\Delta)^{-1}(P,Q)=\\
    =&-N^{-1}(P)\big\{(J(P+\Pi )-\Delta_\rho)^*\delta(P-Q)\\
    &+\Delta_a^*\delta(P+\Pi -Q)\big\}
  \end{split}
\end{equation}
with
\begin{equation}
  \begin{aligned}
    N(P)&=(J(P+\Pi )-\Delta_\rho)^*(J(P)-\Delta_\rho)-\Delta_a^*\Delta_a\\
    &=A(P)+2i\omega_F\mu_\text{eff}\,\,\text{diag}(1,1,-1,-1),\\
    A(P)&=\omega_F^2+h_a^2\vec a^2-\mu_\text{eff}^2+4t^2(\cos p_1+\cos p_2)^2,\\
    \mu_\text{eff}&=\mu+h_\rho\rho.
  \end{aligned}
\end{equation}
Inserting eq. \eqref{eq:inverse_propagator} into eq. \eqref{eq:schwdyslhs} and performing the Fourier transform of eq. \eqref{eq:gapfun_a} (with $g(X)=\sum_Q e^{iQX}g(Q)$) we arrive at the gap equation for $h_a\vec a$
\begin{equation}
  \label{eq:finalgap}
  1=U\sum_Q\frac{A(Q)}{A^2(Q)+4\omega_F^2\mu_\text{eff}^2}.
\end{equation}

\begin{figure}[h!tb]

\centering
\psfrag{mu}{$\mu/t$}
\psfrag{T}{$T/t$}
\includegraphics[scale=0.65]{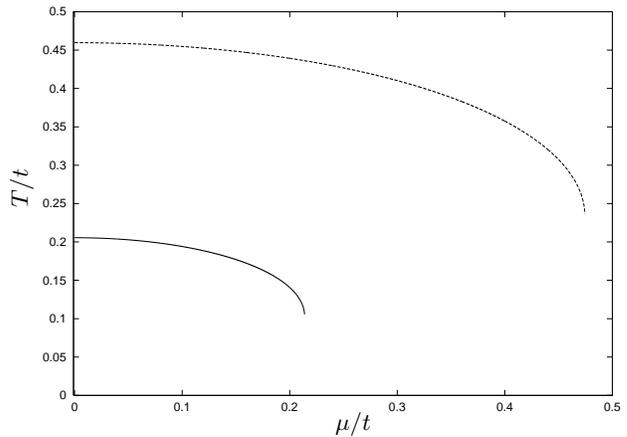}
\caption{Phase boundary for low temperature antiferromagnetism in the
lowest order Schwinger-Dyson and mean field approximations. We show the
critical temperature as a function of the chemical potential $\mu$. We choose
couplings ($U_m=U/3$, $U_\rho=0$) (lower line) and ($U_m=U/2$, $U_\rho=U/2$)
(upper line). The upper line corresponds to the the solution of the Schwinger-Dyson
equation. Both curves correspond to the {\em same} value $U/t=3$ and their difference
reflects the Fierz ambiguity discussed in sect. \ref{sec:fierz}.}
\label{fig:ambiguity}
\end{figure}

Equation \eqref{eq:finalgap} admits a solution with $\vec a \neq 0$ for low enough $T$ and $\mu_\text{eff}$. This nontrivial solution has indeed a lower value of the relevant free energy \cite{Bai01} as compared to the solution with $\vec a=0$. It is therefore a good candidate for the thermal equilibrium state at low $T$. (We will argue in more detail below (sect. VIII) that this state indeed minimizes the appropriate free energy in the approximation of the lowest order Schwinger--Dyson equation.) The onset of spontaneous symmetry breaking ($\vec a\neq 0$) -- as the temperature is lowered from high values where $\vec a=0$ -- can be associated with a phase transition. We have solved eq. \eqref{eq:finalgap} numerically \cite{Bai00} and the result for the critical line in the plane $(T,\mu_\text{eff})$ is shown in fig.\ref{fig:ambiguity} (upper curve).

We can extend the analysis for other possible forms of the local gap (or $g(X)$). For example, ferromagnetism (instead of antiferromagnetism) can be described by replacing in eq. \eqref{eq:gapfun_a} $h_a\vec a \exp(i\Pi X)\to h_m\vec m$. We have found that the solutions with $\vec m \neq 0$ have a higher free energy as compared to $\vec a\neq 0$ and that antiferromagnetism is indeed favoured. Charge density waves (with an additional factor $\exp(i\Pi X)$ in eq. \eqref{eq:gabfun_h}) do not allow for nontrivial solutions. The same is true for local Cooper pairs with $s-$, $p-$ or $d-$symmetry. Nonlocal symmetry breaking expectation values $G_{ab}(X,Y\neq X)$ do not contribute to the lowest order Schwinger--Dyson equation due to the locality of the four fermion interaction. Within the Schwinger--Dyson approach an investigation of realistic nonlocal Cooper pairs which could be responsible for superconductivity therefore requires the inclusion of the higher order terms. (See \cite{Wet02} for an explicit formula of the next order term.) As an alternative, one may realize that the renormalisation flow of the four fermion interaction introduces nonlocalities in the effective four fermion interaction \cite{Hal00} and use the ``renormalized gap equation'' proposed in \cite{Wet02}. For nonlocal four fermion interactions even the lowest order Schwinger--Dyson equation can have nontrivial solutions for nonlocal gaps.

We will go beyond the leading order Schwinger--Dyson equation
by using renormalisation group equations
from section \ref{sec:ERGE} on. Before, it will be instructive to understand the connection between the Schwinger--Dyson equation and mean field theory based on partial bosonisation. This will provide us with an appropriate setting for the renormalisation group equation by which we can explore the region of spontaneous symmetry breaking.

\section{Partial bosonisation of the Hubbard model}
\label{sec:partbos}

Among the most promising approaches to investigate the properties of the Hubbard model are renormalisation group studies. So far, these calculations have been mainly carried out in a purely fermionic formulation \cite{Zan97,Hal00,Hon01}. The onset of spontaneous symmetry breaking then shows up in the divergence of the quartic coupling in certain momentum channels, if one follows the renormalisation group flow to larger length scales. This approach has the limitation that one is not able to follow the flow into the region of broken symmetry. This would need the inclusion of higher multifermion interactions and therefore become extremely complex. We adopt here an alternative approach, where symmetry breaking is encoded in non vanishing expectation values of bosonic fields. Each bosonic field corresponds to one possible symmetry breaking channel. This allows us to keep a simple overview in the realistic case of competing channels. In this work we will work with a bosonic field reflecting antiferromagnetic behaviour. Similarly, bosons corresponding to $s$-, $p$- or $d$-wave superconducting behaviour could be defined for an extended discussion \cite{Bai00}.

The main idea is to rewrite the original Hubbard model with a quartic fermionic interaction as an equivalent Yukawa theory, where the quartic fermionic interaction is replaced by fermion bilinears coupled to the bosons \cite{HubStrat}. As mentioned above, symmetry breaking shows up as non vanishing expectation values of some bosonic fields.
These order parameters can be computed by following
the renormalisation group flow for the bosonic effective potential (free energy density)
into the broken symmetry region. In particular, it is now straightforward to investigate the role of bosonic self interactions which correspond to higher multifermion interactions in the purely fermionic language. Furthermore, the identification of symmetry breaking channels with bosonic excitations (particles) enhances our intuitive understanding of what happens during the flow.

However, this approach, as advantageous it may seem at first sight, is plagued by a coupling ambiguity problem. In the fermionic theory, we face one single quartic coupling $U$, whereas in the partially bosonised theory, we introduce as many Yukawa couplings as we introduce bosonic fields.
These Yukawa couplings are not uniquely determined by the equivalence to the original
fermionic model - different choices of these couplings lead to
precisely the same fermionic model if we integrate out the bosons.
This coupling ambiguity would not give rise to any problems were we able to solve the flow equations exactly.
But for many approximations the results will unphysically depend on the choice of the
couplings. At this point, the Schwinger-Dyson approach presented in the last sections
comes in handy. We will see that at least for a given channel the gap equation corresponds
to the mean field equation with a definite choice of the Yukawa couplings.
This determines at least the rough range for the Yukawa couplings which can serve
as a reasonable starting point for approximations beyond the mean field
approximation. We will discuss in sect. XV how the inclusion of the bosonic fluctuation
effects by the use of the partially bosonized
flow equations reduces drastically the dependence of physical quantities on the precise choice of the
Yukawa couplings.

In this section
we will show how to partially bosonise the Hubbard model for the case where we only take into account a {\em charge density boson} $\rho(X)$ and an {\em spin density boson} $\vec m(X)$. These are the same degrees of freedom we used in the ansatz \eqref{eq:rhogap} and \eqref{eq:agap} for the gap function.
Our bosonisation procedure preserves explicitely the continuous $SU(2)$-spin-symmetry
while it is consistent with the Hartree-Fock result. This settles a widely discussed
\cite{MFH} apparent conflict betwen these two aims.
We begin by defining the fermion bilinears
\begin{equation}
  \begin{aligned}
    \tilde\rho(X)&=\hat\psi^\dagger(X)\hat\psi(X),\\
    \tilde{\vec m}(X)&=\hat\psi^\dagger(X)\vec\sigma\hat\psi(X).
  \end{aligned}
\end{equation}
With these bilinears the fermionic coupling term of the original Hubbard model can be written as
\begin{equation}
  (\hat\psi^\dagger(X)\hat\psi(X))^2=\tilde\rho(X)^2=-\tfrac 13\tilde{\vec m}(X)^2.
\end{equation}
Suppressing the explicit  notation for $X$ the fermionic action in the presence of sources for the bilinears is given by
\begin{equation}\label{eq:37}
  S_F=S_{F,\text{kin}}+\tfrac 12 U(\hat\psi^\dagger\hat\psi)^2-J_\rho\tilde\rho-\vec J_m\tilde{\vec m},
\end{equation}
with $S_{F,\text{kin}}$ the fermion kinetic term (\ref{7}). The partition function reads
\begin{eqnarray}
  \label{eq:partition_ferm}
  Z &=& \int {\cal D}(\psi^*,\psi)\exp \big(-(S_F+S_\eta)\big),\nonumber\\
  S_\eta&=&-\eta^\dagger\psi-\eta^T\psi^*.
\end{eqnarray}
We define the partially bosonised partition function by
\begin{equation}
  \label{eq:partition_bos}
  Z[\eta,\eta^*,J_\rho,\vec J_m] = \int {\cal D}(\psi^*,\psi,\hat\rho,\hat{\vec m})\exp \big( - (S+S_\eta+S_J) \big)
\end{equation}
with
\begin{equation}
  \label{eq:action_bos}
  \begin{gathered}
    S=S_{F,\text{kin}}+\tfrac 12 U_\rho\hat\rho^2+\tfrac 12 U_m\hat{\vec m}^2-U_\rho\hat\rho\tilde\rho-U_m\hat{\vec m}\tilde{\vec m},\\
    S_J = -J_\rho\hat\rho-\vec J_m\hat{\vec m}.
  \end{gathered}
\end{equation}
The functional integral now also involves the bosonic fields $\hat\rho$ and $\hat{\vec m}$. The equations of motion for the bosons yield
\begin{equation}
  \hat\rho=\tilde\rho+\frac{J_\rho}{U_\rho},\qquad\hat{\vec m}=\tilde{\vec m}+\frac{\vec J_m}{U_m}.
\end{equation}
We now demand that inserting the bosonic equations of motion, $S$ should reduce to $S_F$ up to field independent terms. It is then easy to check by Gaussian integration that the partition function \eqref{eq:partition_bos} is equal to \eqref{eq:partition_ferm} up to the logarithm of a field independent quadratic function of the sources \cite{Bai00}. Equivalence between the bosonic and fermionic formulation is therefore achieved for
\begin{equation}
  \label{eq:coupcond}
  U=-U_\rho+3U_m.
\end{equation}
At this point we explicitly see that different combinations of the couplings $U_\rho$ and $U_m$ in the partially bosonised theory lead to the same fermionic theory.

In summary, the fermion--boson action takes the following form in momentum space
\begin{eqnarray}\label{eq:action}
S&=&S_{F,\text{kin}}+S_B+S_Y+S_J,\\
S_{F,\text{kin}}&=&\sum_Q\hat\psi^\dagger(Q)(i\omega_F-\mu-2t(\cos q_1+\cos q_2))\hat\psi(Q),\nonumber\\
S_B&=&\frac{1}{2}\sum_Q\left(U_\rho\hat\rho(Q)\hat\rho(-Q)+U_m\hat{\vec m}(Q)\hat{\vec m}(-Q)\right),\nonumber\\
S_Y&=&-\sum_{QQ'Q''}\delta(Q-Q'+Q'')\times\nonumber\\
&&\hspace{-1cm}\Big(U_\rho\hat\rho(Q)\hat\psi^\dagger(Q')\hat\psi(Q'')+U_m\hat{\vec m}(Q)\hat\psi^\dagger(Q')\vec\sigma\hat\psi(Q'')\Big),\nonumber\\
S_J&=&-\sum_Q\left(J_\rho(-Q)\hat\rho(Q)+\vec J_m(-Q)\hat{\vec m}(Q)\right).\nonumber
\end{eqnarray}
Here the bosonic momenta $Q=(\omega_B,q_1,q_2)$ involve Matsubara frequencies $\omega_B=2\pi mT$ with integer $m$.

\section{A mean field calculation}
\label{sec:MF}

In this section we will explicitly calculate the effective action in an approximation
that sets the bosonic fields to constant background fields. Only the fermion
fluctuations are included in the functional integral (\ref{eq:partition_bos}).
In this ``mean field theory'' (MFT) the fermionic integration can be carried out analytically. In particular, constant bosonic ``background fields'' $\rho$ and $\vec a$
\begin{equation}
  \begin{aligned}
    \hat\rho(Q)&\to\rho\delta(Q)\\
    \hat{\vec m}(Q)&\to\vec a\delta(Q-\Pi),
  \end{aligned}
\end{equation}
correspond to homogeneous charge density ($\rho$) and antiferromagnetism ($\vec a$).
The mean field partition function reads
\begin{eqnarray}
Z_\text{MF}&=&\int{\cal D}(\hat\psi^*,\hat\psi)\exp(-S_\text{MF}),\\
S_\text{MF}&=&\sum_Q\hat\psi^\dagger(Q)(i\omega_F-\mu-2t(\cos q_1+\cos q_2))\hat\psi(Q)\nonumber\\
&&-\sum_Q(U_\rho\rho\hat\psi^\dagger(Q)\hat\psi(Q)+U_m\vec a\hat\psi^\dagger(Q+\Pi)\vec\sigma\hat\psi(Q))\nonumber\\
&&+\frac{V_2}{2T}(U_\rho\rho^2+U_m\vec a^2)-J_\rho(0)\rho-\vec J_m(-\Pi)\vec a,\nonumber
\end{eqnarray}
with $V_2$ the two dimensional volume of the lattice. From the effective action
\begin{equation}
  \Gamma_\text{MF}=-\ln Z_\text{MF}+J_\rho(0)\rho+\vec J_m(-\Pi)\vec a
\end{equation}
we can derive the {\em effective potential}\footnote{There should be no confusion between the effective potential and the coupling in the Hubbard model which traditionally are both denoted by the letter $U$.}
\begin{equation}\label{46AA}
U(\rho,\vec a)=\frac{T\Gamma}{V_2}=\tfrac 12(U_\rho\rho^2+U_m \vec a^2)+\Delta U(\rho,\vec a),\\
\end{equation}
with the fermionic one loop contribution
\begin{eqnarray}\label{eq:U_mf}
&&\Delta U(\rho,\vec a)=-\frac{T}{V_2}\ln\int{\cal D}(\hat\psi^*,\hat\psi)\exp(-S_\Delta),\\
&&S_\Delta=\sum_Q\left(\hat\psi^\dagger(Q)P(Q)\hat\psi(Q)-U_m\vec a\hat\psi^\dagger(Q+\Pi)\vec\sigma\hat\psi(Q)\right),\nonumber
\end{eqnarray}
and $P(Q)$ now given by
\begin{equation}
  \label{46BB}
  \begin{aligned}
    &P(Q)=i\omega_F-\mu_\text{eff}-2t(\cos q_1+\cos q_2),\\
    &\quad\mu_\text{eff}=\mu+U_\rho\rho.
  \end{aligned}
\end{equation}
By defining
\begin{equation}
  \tilde\psi(Q)=\begin{pmatrix}\hat\psi(Q)\\\hat\psi(Q+\Pi)\end{pmatrix}
\end{equation}
we can cast $\Delta U$ into the form
\begin{equation}
  \begin{aligned}
    &\Delta U(\rho,\vec a)=\\
    &=-\frac{T}{V_2}\ln\int{\cal D}(\tilde\psi^*,\tilde\psi)\exp\left(-\frac{1}{2}\sum_Q\tilde\psi^\dagger(Q)\tilde P(Q)\tilde\psi(Q)\right)\\
    &\tilde P(Q)=\begin{pmatrix}P(Q)&-U_m\vec a\vec\sigma\\-U_m\vec a\vec\sigma&P(Q+\Pi)\end{pmatrix}.
  \end{aligned}
\end{equation}
The integral is Gaussian and can be carried out
\begin{eqnarray}\label{eq:mfUnomats}
\Delta U(\rho,\vec a)&=&-\frac{1}{2}\sum_Q\ln\big(-\det\tilde P(Q)\big)\\
&=&-\sum_Q\ln\Big(\omega_F^2-\mu_\text{eff}^2+4t^2(\cos q_1+\cos q_2)^2\nonumber\\
&&\qquad\qquad\qquad+U_m^2\vec a^2+2i\omega_F\mu_\text{eff}\Big).\nonumber
\end{eqnarray}
The Matsubara sum in $\Delta U$ can be analytically performed, so that we finally find (apart from a temperature dependent constant)
\begin{eqnarray}\label{eq:mfUmats}
\Delta U&=&-T\int_{-\pi}^\pi\frac{d^2q}{(2\pi)^2}\sum_{\epsilon\in\{-1,1\}}\ln\cosh\\
&&\bigg(\frac{1}{2T}\Big(\mu_\text{eff}+\epsilon\sqrt{4t^2(\cos q_1+\cos q_2)^2+U_m^2\vec a^2}\Big)\bigg).\nonumber
\end{eqnarray}

In order to establish the connection between the mean field approach and the lowest order Schwinger-Dyson equation we take a closer look at eq. \eqref{eq:mfUnomats}. The value of $\vec a^2$ minimising the effective potential is zero for the symmetric phase, but differs from zero in the case that the the symmetry is spontaneously broken, which in our case means that the system shows antiferromagnetic behaviour. The necessary condition for this is $\frac{\partial U}{\partial\vec a^2}=0$, which can be brought into the form
\begin{equation}
  1=2U_m\sum_K\frac{A(K)}{A^2(K)+4\omega_F^2\mu_\text{eff}^2}.
\end{equation}
This is the same as eq. \eqref{eq:finalgap}, if we identify the gap $h_\rho\rho$ in the gap equation (sect. \ref{sec:antifer_gap}) with $U_\rho\rho$ here (so that $\mu_\text{eff}$ used in the gap equation and the mean field calculation coincide) and similarly $U_m^2\vec a^2=h_a^2\vec a^2$. We also have to set $U_m=U/2$, which implies $U_\rho=U/2$ by eq. \eqref{eq:coupcond}. The Schwinger-Dyson approach therefore does not have the coupling ambiguity introduced in the partial bosonisation procedure.
As mentioned before, it corresponds to a special choice of the couplings in mean field theory.

\section{Mean field ambiguity}
\label{sec:fierz}

For high enough temperature and small enough coupling an expansion in the small dimensionless quantity $U/T$ is expected to be valid. In this limit the lowest order Schwinger--Dyson equation becomes the lowest order in a systematic expansion where higher orders are suppressed by higher orders of $U/T$. (This is not a simple expansion in powers of $U/T$ like perturbation theory but rather corresponds to a resummed expansion.) For small $U/T$ mean field theory therefore gives an incorrect result unless we choose the particular bosonisation procedure with $U_m=U_\rho=U/2$. Since the functional integral transformations leading to partial bosonisation are exact the failure of mean field theory for other choices of $U_m$ and $U_\rho$ must be due to the bosonic fluctuations which have not yet been included. If we were able to compute the effects of the bosonic fluctuations exactly the final results would have to be completely independent of the choice of $U_m$ provided eq. \eqref{eq:coupcond} is obeyed. In general, the inclusion of the bosonic fluctuations is therefore crucial for a reliable picture. This issue has been systematically studied in \cite{Jae02}. For small $U/T$ it just happens that the sign of the effect of the bosonic fluctuations may be positive or negative, depending on the value of $U_m$. In particular, we infer that to lowest order in $U/T$ the bosonic correction to the (anti)ferromagnetic correlation function $\langle\tilde m(X)\tilde m(Y)\rangle$ precisely vanishes for $U_m=U/2$.

Near the phase transition an expansion in small $U$ is no longer reliable, the characteristic quantity $U/|T-T_c|$ even diverges for $T\to T_c$. The corrections to the lowest order Schwinger--Dyson equation become important. In consequence, there is no reason to believe that the bosonic fluctuations beyond mean field theory are a small effect for $U_m=U/2$. We conclude that for the range of large $U/|T-T_c|$ the ``gap equation'' value $U_m=U/2$ is at best a good educated guess. We feel that it is much safer to study a whole range\footnote{One  should, however, avoid the extreme scenarios where $U_m/U$ becomes very large or very small or negative. A reasonable range may be within 30\% of $U_m/U=0.5$.} of choices for $U_m$. The dependence on the choice of $U_m/U$ in mean field theory may then be turned into an advantage when fluctuations beyond mean field theory are included: if an approximation beyond mean field theory is valid the dependence on the unphysical parameter $U_m/U$ should get considerably reduced as compared to mean field theory. This becomes a demanding check for the validity of approximations. We will see in section \ref{sec:remove_ambig} that our renormalisation group approach indeed has this important property. For the moment, we illustrate the mean field ambiguity by the lower curve in the phase diagram in fig. \ref{fig:ambiguity} which corresponds to the choice $U_m=U/3$, $U_\rho=0$.
In the language of the present section this figure shows the mean field results for the onset of spontaneous symmetry breaking towards the antiferromagnetic state for
two choices of the Yukawa couplings $U_\rho,U_m$, namely
($U_m=U/3$, $U_\rho=0$) (lower line) and for ($U_m=U/2$, $U_\rho=U/2$) (upper line). The lines indicate a second order phase transition and stop where the phase transition becomes of first order. We observe that without the inclusion of the bosonic fluctuations the ``mean field ambiguity'' for the phase diagram is considerable. Depending on the choice of $U_m/U$ the critical temperature varies by more than a factor of two,
limiting severely the reliability of mean field theory.

\section{Mean field calculation of the bosonic propagator}
\label{sec:bosprop}

In this section we will calculate the bosonic propagator for $\hat{\vec m}$ in mean field theory. This also gives the bosonic propagator according to the lowest order Schwinger--Dyson equation if we choose $U_m=U/2$. There are two reasons why we are interested in the result. In the derivation of the gap equation and the mean field approximation, we have {\em assumed} that antiferromagnetism is the dominating mechanism by which the $SU(2)$-symmetry is spontaneously broken. By inspection of the bosonic propagator as a function of momentum we will see that indeed antiferromagnetism (in contrast to e.g. ferromagnetism) is the favoured mode of symmetry breaking. The second reason is that --- as we will see in the subsequent sections --- our renormalisation group equations take the form of one loop equations. The calculation in this section therefore can be immediately used as a technical ingredient for the derivation of the renormalisation group equations.
Furthermore, the general features of the momentum dependence of the bosonic propagators
found in this section will be useful in order to device truncations for the flow equations.

The bosonic propagators are propagators for fermion bilinears and therefore correspond to particular combinations of four fermion correlation functions in the original fermionic description. They are particularly simple to describe in the partially bosonised language of section \ref{sec:partbos}. For example, the propagator for the spin wave boson
\begin{equation}
  \label{eq:V1}
  \langle m_i(X)m_j(Y)\rangle_c = (G^m_B)_{ij}(X,Y)
\end{equation}
is given by the inverse of the second functional derivative of the effective action
with respect to the scalar fields
in the appropriate channel. Here the effective action in the partially bosonised language obtains from the partition function \eqref{eq:partition_bos} by the usual Legendre transform
\begin{equation}
  \label{eq:V2}
  \Gamma[\rho,\vec m,\psi,\psi^*] = -\ln Z + J_\rho\rho + \vec J_m \vec m + \eta^\dagger\psi + \eta^T\psi^*.
\end{equation}
As usual the second functional derivative of $\Gamma$ with respect to $\rho$ and
$\vec{m}$ is directly related to the second functional derivative of $\ln Z$ with respect
to the sources $J_\rho$ and $J_m$ (by inversion). Applying this to eqs. (\ref{eq:partition_ferm}) and
(\ref{eq:37}) and setting $J_\rho=J_m=0$ we see indeed how particular combinations of
fermionic four point functions can directly be read off from the bosonic inverse
propagator \cite{Bai01}.

In absence of spontaneous spin--symmetry breaking the inverse bosonic propagator for $\vec m$ obtains by expanding $\Gamma$ in second  order in $\vec m$ for fixed constant $\rho$ and $\psi=\psi^*=0$
\begin{gather}
  \label{eq:V3}
  \Gamma = \frac 12 \sum_{QQ'} m_i(-Q)(\Gamma^{(2)}_m)_{ij}(Q,Q') m_j(Q') + \cdots,\\
  \Gamma^{(2)}_m(Q,Q') = \hat\Gamma^{(2)}_m(Q)\delta(Q-Q')
\end{gather}
(with $G^{(m)}_B=(\Gamma^{(2)}_m)^{-1}$). In section VI we have already computed $\hat\Gamma^{(2)}_m(Q=\Pi)$ in equations \eqref{eq:U_mf}, \eqref{eq:mfUnomats}. It corresponds to the antiferromagnetic susceptibility or inverse correlation length
\begin{equation}
  \label{eq:V4}
  \begin{gathered}
    (\hat\Gamma^{(2)}_m)_{ij}(\Pi) = \bar m_a^2 \delta_{ij},\qquad
    \bar m_a^2 = U_m + 2\frac{\partial}{\partial (\vec a^2)} \Delta U|_{\vec a^2=0}.
  \end{gathered}
\end{equation}
The critical line for the phase transition shown in fig. \ref{fig:ambiguity} corresponds to $\bar m_a^2(T=T_c)=0$. If we reexpress this in the fermionic language the four fermion interaction in the antiferromagnetic channel diverges for $T\to T_c$ proportional to $\bar m_a^{-2}$.

In this section we want to compute $\hat\Gamma^{(2)}_m(Q)$ for arbitrary $Q$. Antiferromagnetism is the preferred condensate (with lowest free energy) if $\hat\Gamma^{(2)}_m(Q)$ has its minimum for $Q=\Pi$. (A minimum at $Q=0$ would favour a ferromagnetic condensate.) We start from the action \eqref{eq:action} and treat the bosonic fields as background fields according to
\begin{equation}
  \hat\rho(Q)\to\rho, \quad \hat{\vec m}(Q)\to\vec m(Q).
\end{equation}
Note that we now keep the momentum dependence of $\vec m$ to be able to distinguish between ferromagnetic, antiferromagnetic, etc. behaviour. For a (mean field) computation of the fermionic fluctuation effects we need the expansion of the action (\ref{eq:action}) in second order in the fermion fields
\begin{equation}
S_2=\sum_{QQ'}\hat{\psi}^\dagger(Q)(P_0(Q,Q')+\Delta P(Q,Q'))\hat{\psi}(Q')
\end{equation}
with
\begin{equation}
  \begin{split}
    P_0(Q,Q')&=P(Q)\delta(Q-Q')\\
    \Delta P(Q,Q')&=-U_m\vec m(Q-Q')\vec\sigma
  \end{split}
\end{equation}
and $P(Q)$ given by eq. (\ref{eq:action}) in second order in the fermion fields
The fermionic functional integral can be carried out and we get
\begin{eqnarray}\label{zmf_graph_bosexpand}
&&\Delta\Gamma = -\ln\int{\cal D}(\hat{\psi}^*,\hat{\psi})\exp(-S_2) \\
&&=-\ln\det(P_0+\Delta P)\nonumber\\
&&=\Delta\Gamma_0+\Delta\Gamma_1+\Delta\Gamma_2+\cdots\nonumber\\
&&=-\Tr\ln P_0-\Tr(P_0^{-1}\Delta P)+\frac{1}{2}\Tr(P_0^{-1}\Delta P P_0^{-1}\Delta P).\nonumber
\end{eqnarray}
The first term is independent of $\vec m$, the second vanishes, and $\Delta\Gamma_2$ is the $\vec m$-propagator correction we are looking for. Its second derivative is given by
\begin{eqnarray}\label{eq:bos_prop_korr}
(\Delta\hat\Gamma^{(2)}_m)_{ij}(Q) &=& \Delta P_m(Q) \delta_{ij}\\
\Delta P_m(Q) &=& 2U_m^2\sum_KP^{-1}(K)P^{-1}(K+Q).\nonumber
\end{eqnarray}
If we set $Q=\Pi$, we reproduce the term in the mean field result of quadratic order in $\vec a$.

The inverse propagator for the spin wave bosons can now be computed numerically after adding the classical piece \eqref{eq:action_bos} $S_0$, i.e. $P_m(\omega_B,q_1,q_2)=\hat\Gamma^{(2)}_m(Q) = U_m+\Delta P_m(\omega_B,q_1,q_2)$. For vanishing Matsubara frequency (i.e. $Q=(0,q_1,q_2)$) and for $\mu=0$ the Matsubara sums in \eqref{eq:bos_prop_korr} can be performed analytically
\begin{equation}
  \Delta P_m(Q) = U_m \frac{2T}{(\pi T)^2}\int_{-\pi}^\pi\frac{d^2q}{(2\pi)^2}S(m, \frac{\epsilon_{\Bs q}}{\pi T}, \frac{\epsilon_{\Bs q+\Bs k}}{\pi T}).
\end{equation}
For this purpose we use the identity
\begin{eqnarray}
  S(m;a,b) &=&
  \sum_{n\in \mathbbm{Z}}\frac{ab-(2n+1)(2(n+m)+1)}{[a^2+(2n+1)^2][b^2+(2(n+m)+1)^2]} \nonumber\\
  &=& -\frac{\pi}{2} \frac{(a-b)(\tanh\tfrac{a\pi}{2}-\tanh\tfrac{b\pi}{2})}{4m^2+(a-b)^2},
\end{eqnarray}
which has the following limits
\begin{equation}
  \begin{split}
    S(0;a,a)  &= -\frac{\pi^2}{4} \cosh^{-2}\tfrac{a\pi}{2},\\
    S(0;a,-a) &= -\frac{\pi}{2}\frac{\tanh\tfrac{a\pi}{2}}{a}.
  \end{split}
\end{equation}

\begin{figure}[tb]

\psfrag{k1}{\footnotesize $q_1$}
\psfrag{k2}{\hspace{-4mm}\footnotesize $q_2$}
\psfrag{P}{\hspace{-4mm}\footnotesize $P_m$}
\hspace{-0.5cm}
\includegraphics[scale=0.7]{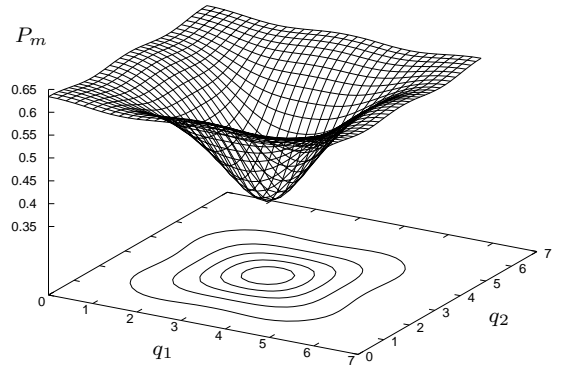}
\hspace{-1.5cm}
\includegraphics[scale=0.7]{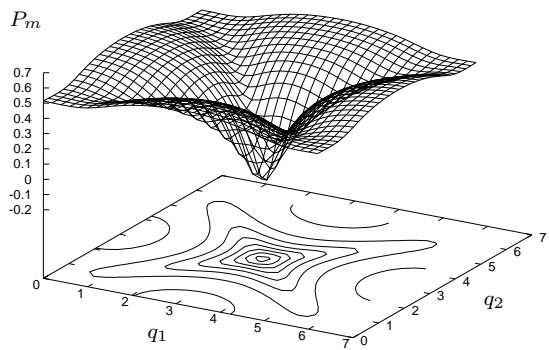}

\caption{Mean field approximation for the inverse propagator of spin waves as a function of momenta. We show two different temperatures $T=0.5t$ (upper part) and $T=0.15t$ (lower part) for $\mu_\text{eff}=0$, $\omega_B=0$ and choose $U_m=t$.}
\label{fig:oneloop_prop_m_k}
\end{figure}

We show the momentum dependence of $P_m$ in fig. \ref{fig:oneloop_prop_m_k} and observe a pronounced minimum for $Q=\Pi$. This clearly establishes that antiferromagnetism is the preferred condensate as compared to other spin waves with $Q\neq \Pi$. We also note that the minimum of $P_m$ gets more pronounced as the temperature is lowered, reaching $P_m(\Pi)=0$ for $T=T_c$. (For our example $T_c\approx 0.2t$.) Also notice the development of sharp crests at low temperature (lower figure) due to the singularities in the fermionic propagators at the Fermi surface.

\begin{figure}[tb]
\psfrag{P}{\hspace{-4mm}\footnotesize $P_m$}
\psfrag{m}{\hspace{-1mm}\footnotesize $m$}
\hspace{-0.4cm}
\includegraphics[scale=0.6]{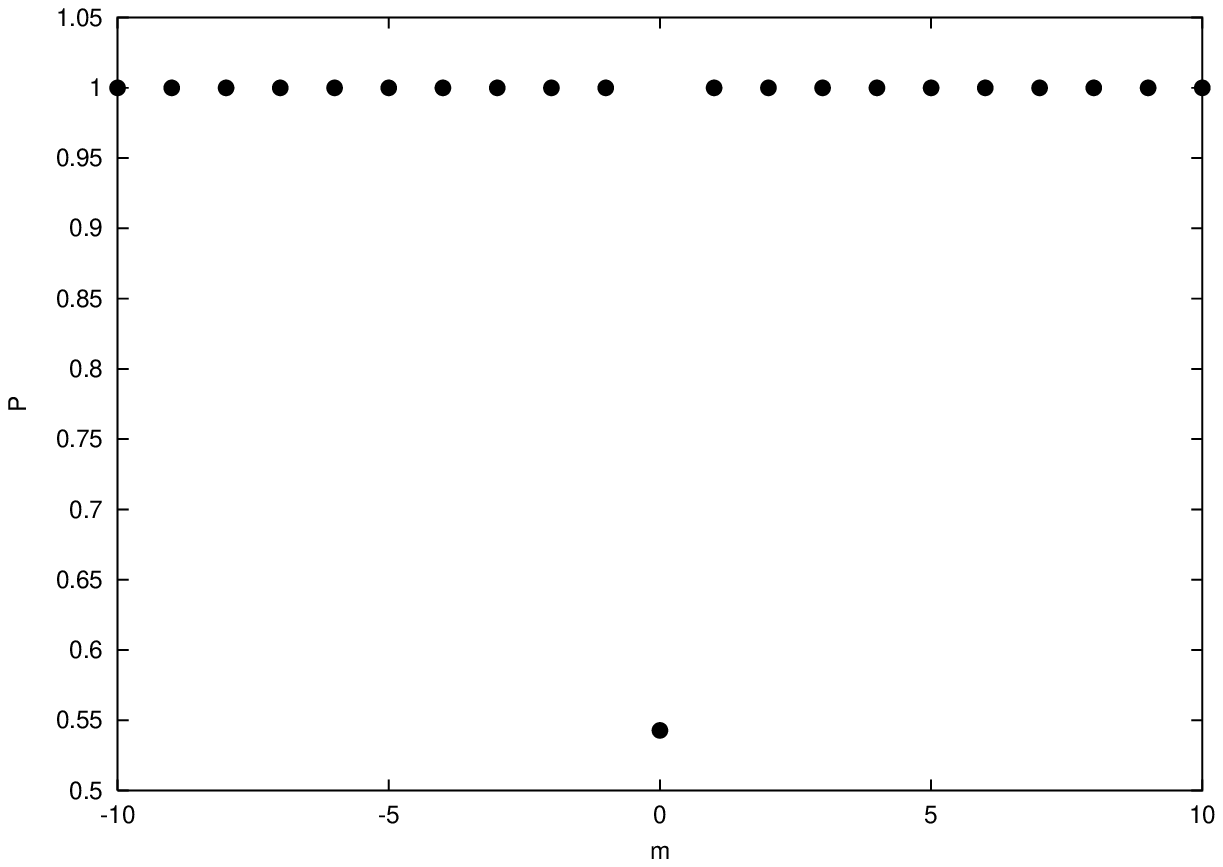}
\includegraphics[scale=0.6]{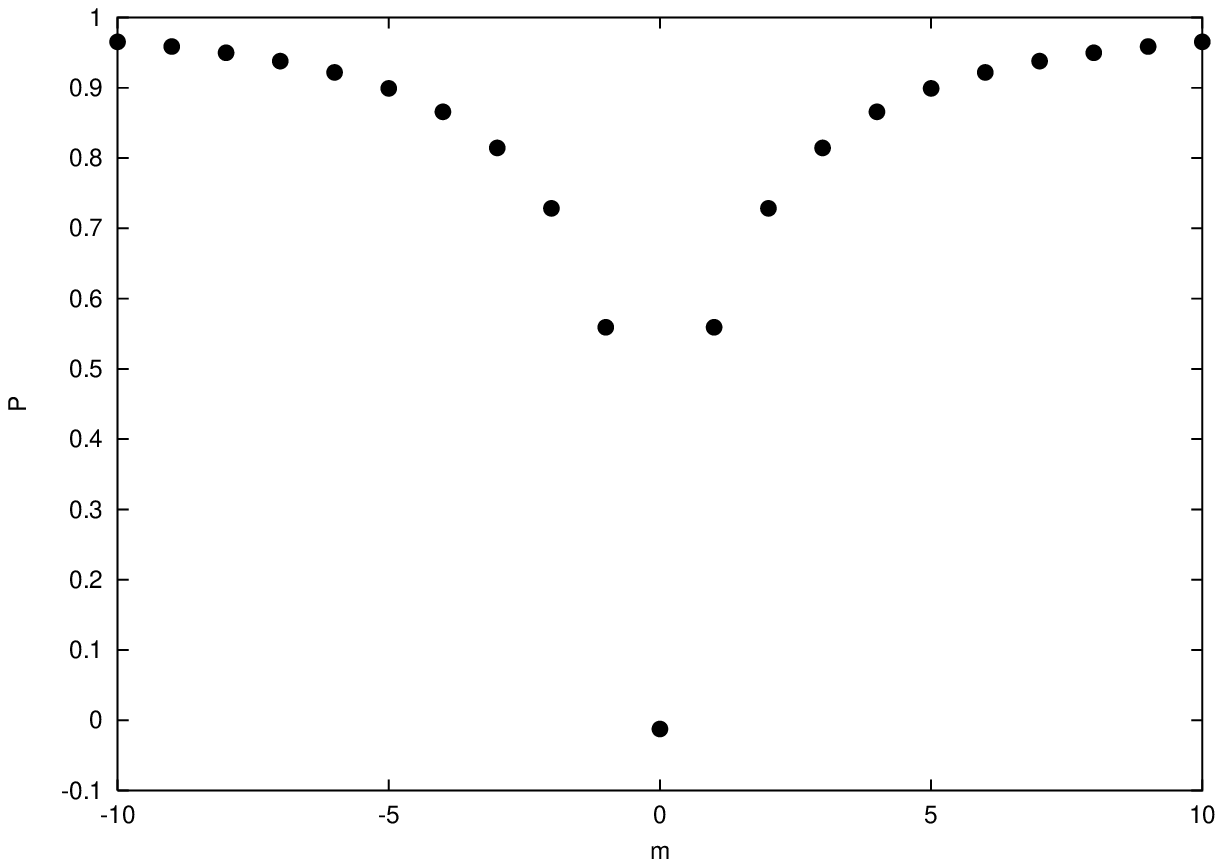}

\caption{Inverse propagator for the boson $\vec m$ as a function of the Matsubara frequency $\omega_B=2\pi mT$ at $T=0.2t$ for $U_m=t$. We show $q_1=q_2 = 0$ (upper part) and $q_1=q_2 = \pi$ (lower part).}
\label{fig:oneloop_prop_m_mats}
\end{figure}

Fig. \ref{fig:oneloop_prop_m_mats} shows the dependence of $P_m(Q)$ on the Matsubara frequency $\omega_B=2\pi mT$ for two values of the external momenta. Note that the $\omega_B = 0$ mode is the one that is changed most by fluctuation effects. For $q_1=q_2 = 0$ this is even the only mode that is changed at all, while $\Delta P_m$ vanishes in the other cases. We observe that $P_m$ is even in $\omega_B$ as requested by the discrete (time reversal) symmetry of the Hubbard model.

A main reason for calculating the one loop corrections to the bosonic propagators was to get a feeling for the momentum dependence the propagators are likely to obtain under a renormalisation group flow. In sect. XI we will make use of these observations in order to formulate suitable truncations for the bosonic propagators. A more detailed discussion of the propagators of the other bosonic fields (charge density waves, Cooper pairs) can be found in \cite{Bai01}.

\section{Exact renormalisation group equation}
\label{sec:ERGE}

In the remainder of this paper we want to extend our analysis of the Hubbard model beyond the approximation of mean field theory. Our tool will be nonperturbative flow equations for the scale dependence of the effective average action \cite{Wet91}. The flow equations are obtained from a truncation of the exact functional renormalisation group equation (ERGE) \cite{Wet93a,Ber02}. Within the fermionic language a truncation to momentum dependent four fermion interactions for the ERGE of ref. \cite{Wet93a} has been performed\footnote{Ref. \cite{Hal00} uses a somewhat different formulation of the exact renormalisation group equation.} in \cite{Hon01}. This analysis shows for $T\to T_c$ and $\mu=0$ a leading divergence of the four fermion interaction in the antiferromagnetic channel. It also covers the issue of superconductivity for large enough $|\mu|$ and the analysis has been extended beyond next neighbour coupling.

The limitation of the fermionic language concerns the region when some of the couplings grow very large. One expects that large quartic
fermion couplings also induce large eight fermion couplings and so on. In particular, it seems very difficult to explore the low temperature phase with spontaneous symmetry breaking in this way. For this reason we explore here the formulation in the partially bosonised language of section \ref{sec:partbos}. As we have argued, the divergence of a four fermion coupling just corresponds to the vanishing of the mass of the bosonic field. If higher order bosonic interactions are included in the effective bosonic potential $U$ one can easily explore the region where the curvature of $U$ at the origin gets negative ($m_a^2<0$). This region is characterised by ``spontaneous symmetry breaking''.

Let us consider a theory containing complex bosonic fields $\hat u$, $\hat u^*$, real bosonic fields $\hat w$ and fermionic fields $\hat\psi$, $\hat\psi^*$. We collect the fields into generalised fields and define generalised sources for them (the indices $\alpha$ run over field type, momentum, internal indices etc.)
\begin{eqnarray}
&&\hat\chi_\alpha = (\hat u, \hat u^*, \hat w, \hat\psi, \hat\psi^*)_\alpha,\\
&&J_\alpha = (j^*, j, l, \eta^*, \eta)_\alpha, \nonumber\\
&&S_J[\hat\chi] = -J_\alpha \hat\chi_\alpha = -(j^*\hat u
+ j\hat u^* + l\hat w + \eta^*\hat\psi +\eta \hat\psi^*).\nonumber
\end{eqnarray}

Now we regularise the theory by adding an infrared cutoff $\sim R_k$ to the original action
\begin{gather}
  \begin{aligned}
    \Delta S_k[\hat\chi] &= \tfrac 12 \hat\chi_\alpha R_{k,\alpha\beta}\hat\chi_\beta
    = \hat u^* R_k^u \hat u + \tfrac 12 \hat w^* R_k^w \hat w + \hat\psi^* R_k^\psi \hat\psi.
  \end{aligned}
\end{gather}
It will have the effect of cutting off momenta with $q^2<k^2$ (bosonic fields) or in a shell of size $k$ around the Fermi surface (fermionic fields). In presence of the cutoff the generating functional for the connected Green functions depends on the scale $k$.
\begin{equation}
  \label{eq:defWk}
  W_k[J] = \ln\int{\cal D}\hat\chi \exp\big(-(S[\hat\chi] + \Delta S_k[\hat\chi]) + J\hat\chi)\big).
\end{equation}
More precisely, the function $R_k^\psi(Q)$ is tailored in order to regularise the fermionic zero modes of the propagator. For momenta far from the Fermi surface (compared to $k$) $R_k^\psi$ is to vanish rapidly so that the behaviour of these modes is essentially unaltered. A similar task is assigned to the bosonic cutoff functions. Here the zero modes can occur for $Q^2=0$ on the critical line and $R_k^u(Q)$, $R_k^w(Q)$ act essentially as additional mass terms $\sim k^2$ for the low momentum region. This is easily generalised to the case when the zero mode occurs for some other momentum, e.g. $Q=\Pi$. In the limit $k\to 0$ we demand that the regulators vanish whereas for $k\to\infty$ we assume them to diverge
\begin{equation}
  \label{eq:demandsR}
  \lim_{k\to 0}R_k^\chi = 0,\qquad \lim_{k\to \infty}R_k^\chi = \infty.
\end{equation}
Our specific choice of the cutoff functions will be given in sect. \ref{sec:trunc_antiferr}.

We may now proceed to define the effective average action $\Gamma_k$ in analogy to the definition \eqref{eq:defEA}. By a Legendre transform with respect to the classical fields
\begin{equation}\label{eq:fieldeqAEA}
  \chi = \langle \hat\chi \rangle = \frac{\delta}{\delta J} W_k[J],
\end{equation}
we obtain the functional
\begin{equation}
  \tilde\Gamma_k[\chi] = J\chi - W_k[J],
\end{equation}
where $J=J[\chi]$ is a solution of the equation \eqref{eq:fieldeqAEA}. As will become clear in a moment it is favourable to subtract the cutoff action from this functional and define the {\em effective average action} as
\begin{equation}
  \label{eq:def_AEA}
  \Gamma_k[\chi] = J\chi-W_k[J]-\Delta S_k[\chi].
\end{equation}
The effective average action is the effective action of a theory containing an extra regulator or ``mass'' term as described by the action $S[\hat\chi]+\Delta S_k[\hat\chi]$. Since the effective action respects all (linearly realized symmetries of the original action, this also applies to $\Gamma_k[\chi]$ for all $k$, if the regulator $\Delta S_k[\hat\chi]$ respects the symmetries. It is thus possible to expand the effective average action in invariants with respect to these symmetries.

The limits \eqref{eq:demandsR} lead to corresponding limits\footnote{For $R_k\to\infty$ the fluctuation contribution vanishes (up to a constant).} for the effective average action \cite{Wet93a}
\begin{equation}
  \label{eq:limitsAEA}
  \lim_{k\to 0}\Gamma_k[\chi] = \Gamma[\chi],\qquad \lim_{k\to \infty}\Gamma_k[\chi] = S[\chi].
\end{equation}
This is why we chose to subtract the regulator in the definition of $\Gamma_k[\chi]$: for large ``cutoff'' $k$ this functional is nothing but the original action. The effective average action therefore interpolates smoothly between the classical action $S$ and the effective action $\Gamma$ as $k$ changes from very large values to zero. The quantitative change of $\Gamma_k$ with $k$ will be described by the ERGE derived below.

We specify the second functional derivative in a symmetric form containing both left and right derivatives
\begin{equation}
  F^{(2)}_{\alpha\beta}[\chi] = \tfrac{\stackrel{\rightarrow}{\delta}}{\delta\chi_\alpha} F[\chi]
  \tfrac{\stackrel{\leftarrow}{\delta}}{\delta\chi_\beta}
\end{equation}
and notice the identity
\begin{equation}
  \label{eq:inverseAEA}
  \tilde \Gamma_{k,\alpha\beta}^{(2)} W_{k,\beta\gamma}^{(2)} = M_{\alpha\gamma},
  \qquad M=\text{diag}(1,1,1,-1,-1).
\end{equation}
For the $k$--derivative of $\tilde\Gamma_k$ one now obtains \cite{Wet93a,Ber02}:
\begin{eqnarray}
\partial_k \tilde\Gamma_k[\chi]|_\chi
&&= -\partial_k W_k[J]|_J = \langle \partial_k \Delta S_k \rangle\nonumber\\
&&= \tfrac 12 \partial_k R_{k,\alpha\beta} \langle \chi_\alpha \chi_\beta \rangle\nonumber \\
&&= \tfrac 12 \partial_k R_{k,\alpha\beta}
\big\{ W^{(2)}_{k,\beta\alpha} + \langle\chi_\alpha\rangle \langle\chi_\beta\rangle \big\}
\nonumber \\
&&= \tfrac 12 \partial_k R_{k,\alpha\beta} W^{(2)}_{k,\beta\alpha}
+ \partial_k \Delta S_k[\chi],
\end{eqnarray}
where we used the fact that $W_k[J]$ is the generating functional of connected Green functions. With the aid of \eqref{eq:inverseAEA} we immediately obtain a flow equation for the effective average action
\begin{equation}
  \label{eq:flowAEA}
  \begin{split}
    \partial_k \Gamma_k[\chi]
    &= \tfrac 12 \partial_k R_{k,\alpha\beta} \big[ \Gamma^{(2)}_k+R_k \big]^{-1}_{\beta\gamma}M_{\gamma\alpha}\\
    &= \tfrac 12 \STr\big\{ \partial_k R_k [\Gamma^{(2)}_k+R_k]^{-1} \big\}.
  \end{split}
\end{equation}
Here the ``supertrace'' runs over field type, momentum, internal indices etc. and has an additional minus sign for fermionic entries.

This equation is exact -- we have only performed well defined formal manipulations\footnote{If needed, one may work in a finite volume such that the momentum sum is discrete.}. However, it is an equation for an infinite number of couplings and hence by no means accessible to an exact solution. The usefulness of eq. \eqref{eq:flowAEA} will only show up if we are able to make sensible approximations to the flow equation. We will come back to this later. Here we note that in the flow equation \eqref{eq:flowAEA} the regulator function $R_k$ appears in the effective propagator as an infrared (or Fermi surface) regulator whereas the factor $\partial_kR_k$
reflects the fact that the only $k$-dependence arises through the regulator.
For an appropriate choice of $R_k$ with $\partial_k R_k$ decreasing rapidly for large momenta
the momentum integrals are ultraviolet finite. This
means that effectively only a small interval of momenta contributes to the integrals.

Let us rewrite the flow equation in a very useful way making contact to perturbation theory. Define the derivative (the index $i$ counts the field types)
\begin{equation}
  \tilde\partial_k = (\partial_k R^i_{k})\frac{\partial}{\partial R^i_{k}}.
\end{equation}
With the aid of this derivative the flow equation can be cast into the form
\begin{equation}
  \label{eq:flowAEAtilde}
  \partial_k \Gamma_k[\chi] = \tfrac 12 \STr\big\{ \tilde\partial_k \ln[\Gamma^{(2)}_k+R_k] \big\}.
\end{equation}
This can to be compared with the regularised perturbative one loop result
\footnote{Since we have regularised the propagators we may send $\Lambda$ to very large values.}
\begin{equation}
  \label{eq:perturbative_ERGE}
  \Gamma_k[\chi] = \Gamma_\Lambda[\chi] + \tfrac 12 \STr\big(\ln[S^{(2)}+R_k]-\ln[S^{(2)}+R_\Lambda]\big).
\end{equation}
Performing the $k$--derivative of eq. \eqref{eq:perturbative_ERGE} leads to a one loop flow equation. A ``renormalisation group improvement'' $S^{(2)}\to\Gamma^{(2)}_k$ promotes this equation to a non--perturbative exact flow equation \cite{Wet91}. This comparison with perturbation theory allows us to identify the right hand side of \eqref{eq:flowAEAtilde} as a sum of one particle irreducible one loop diagrams, where all couplings have been replaced by their renormalised counterparts. Momentum integrations, sums over internal indices etc. are performed after the $\tilde\partial_k$ derivative.

Obtaining the flow equation for some coupling thus amounts to summing all one loop diagrams for this coupling, evaluating  the $\tilde\partial_k$ derivative and then calculating the trace. However, we may be able to perform parts of the trace first if the cutoff does not depend on it. For example we will later be able to first sum over Matsubara indices before performing the $\tilde\partial_k$ derivative.
Technically, this is a very useful property since it allows us to use directly the one loop computations (e. g. sect. VIII) for a derivation of the appropriate renormalization flow equations.

On the level of the exact flow equation we have, of course, to remember that the one loop diagrams have to be evaluated for field dependent propagators as given by $(\Gamma_k^{(2)}+R_k)^{-1}$. Therefore the flow equation \eqref{eq:flowAEA} is a complex differential equation for functionals. In this paper we tackle it by expanding the effective action in powers of the fields
\begin{equation}
  \Gamma_k[\chi] = \sum_{n=0}^\infty \frac{1}{n!} \sum_{\alpha_i}
  \chi_{\alpha_1}\cdots\chi_{\alpha_n} \Gamma^{(n)}_{k;\alpha_1,\ldots,\alpha_n}.
\end{equation}
The flow equations of the $n$--point functions $\Gamma_k^{(n)}$ can easily be derived from eq. \eqref{eq:flowAEA} by appropriate functional derivatives. However, the flow of some $n$--point function will in general contain higher $n$--point functions. This is a general feature: if we perform a systematic expansion of the effective average action, the set of flow equations will not be closed. We have to truncate the expansion at some point.

\section{Rebosonisation of four fermion interactions}
\label{sec:rebos}

Any partially bosonised theory will generate four fermion interaction terms under a
renormalisation group step corresponding to the diagrams in Fig. 5.
\begin{figure}[h]
\includegraphics[width=6cm]{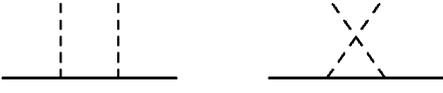}
\caption{Box diagrams for four fermion interactions. Solid lines denote fermions, dashed lines bosons.}
\end{figure}

\noindent
However, we wanted to capture the complicated behaviour of higher fermion vertices in the bosonic language -- this is what the bosonisation procedure is all about. One might suspect that it should be possible to rebosonise the fermionic coupling obtained after some renormalisation group step by a suitable field redefinition of the bosonic fields. This is indeed the case as was shown in \cite{Gie02} (see also \cite{Jae02}).

Consider a theory with (effective average) action

\begin{equation}
  \label{eq:trunc_lambda}
  \begin{split}
    \Gamma_k[\psi,\psi^*,\phi] =& \sum_Q \psi^*(Q) P_{\psi,k} \psi(Q)\\
    & + \tfrac 12 \sum_Q \phi(-Q) P_{\phi,k}(Q) \phi(Q)\\
    & - \sum_Q h_k(Q) \phi(Q) \tilde\phi(-Q)\\
    & + \sum_Q \lambda_{\psi,k}(Q) \tilde\phi(Q)\tilde\phi(-Q),
  \end{split}
\end{equation}
where $\tilde\phi$ is the fermionic bilinear corresponding to the bosonic field $\phi$, e.g.
$\tilde\phi_i(Q) = \sum_K \psi^*(K)\sigma^i\psi(Q+K)$,
and the initial condition for the purely fermionic coupling is $\lambda_{\psi,\bar k}=0$ at some initial scale $\bar k$.

Now perform a renormalisation group step from the scale $\bar k$ to the scale $k=\bar k-\Delta k$. The change in scale, $\Delta k$, is supposed to be so small that the changes in couplings are also small; they are calculated by the flow equation \eqref{eq:flowAEA} for the truncation \eqref{eq:trunc_lambda}. As can be seen from the diagrams in Fig. 5, the four fermion coupling $\lambda_{\psi,k}$ will in general be different from zero, say $\Delta \lambda_{\psi,k}$, at the new scale $k$.

We will use our freedom in the definition of the bosonic fields to consider a field redefinition at the scale $k$
\begin{equation}\label{eq:fieldredefinition}
\phi_k(Q) = \phi_{\bar k}(Q) + \Delta \alpha_k (Q)\tilde\phi(Q),
\end{equation}
where $\Delta \alpha_k$ is an up to now arbitrary function. (We put $\phi_{\bar k} = \phi$ at the initial scale.) For infinitesimally small shifts this results in a flow equation for a $k$--dependent field variable $\phi_k$
\begin{equation}
  \partial_k \phi_k(Q) = - \, \partial_k\alpha_k(Q)\tilde\phi(Q).
\end{equation}
Let us see how we can implement the field redefinition into the renormalisation group formalism. In the exact renormalisation group equation \eqref{eq:flowAEA} the change of scale $\partial_k\Gamma_k[\chi]|_\chi$ is calculated at fixed fields. Hence, if in addition we perform a shift in the fields as above and consider the flow for fixed $\phi_k$, the flow equation becomes
\begin{equation}\label{86}
  \begin{split}
    \partial_k\Gamma_k&[\psi,\psi^*,\phi_k] = \partial_k\Gamma_k[\psi,\psi^*,\phi_k]|_{\phi_k} \\
    &\qquad\qquad\qquad + \sum_Q \big( \tfrac{\delta}{\delta \phi_k}\Gamma[\psi,\psi^*,\phi_k] \big)\partial_k\phi_k \\
    =& \partial_k\Gamma_k[\psi,\psi^*,\phi_k]|_{\phi_k} \\
    &\quad+ \sum_Q \big(-\partial_k\alpha_k(Q)P_{\phi,k}(Q) \phi_k(Q)\tilde\phi(-Q) \\
    &\qquad\qquad\qquad+h_k(Q)\partial_k\alpha_k(Q) \tilde\phi(Q)\tilde\phi(-Q) \big).
  \end{split}
\end{equation}
It is appropriate to define the couplings $h_k$ etc. as the coefficients of polynomials in $\phi_k$. The secons term in eq. (\ref{86}) this changes the flow equations for $h_k$ and $\lambda_{\psi,k}$ to
\begin{equation}
  \begin{split}
    \partial_k h_k(Q) &= \partial_k h_k(Q)|_{\phi_k} + \partial_k \alpha_k(Q) P_{\phi,k}(Q), \\
    \partial_k \lambda_{\psi,k}(Q) &= \partial_k\lambda_{\psi,k}(Q)|_{\phi_k} + h_k(Q)\;\partial_k\alpha_k(Q).
  \end{split}
\end{equation}
We may now demand that the purely fermionic coupling vanishes for all scales $k$, i.e. $\partial_k \lambda_{\psi,k}(Q)\equiv 0$. This determines $\partial_k \alpha_k(Q)$ and leads to a modified flow equation for the Yukawa coupling
\begin{equation}
  \label{eq:rebos_corr}
  \partial_k h_k(Q) = \partial_k h_k(Q)|_{\phi_k} - \frac{P_{\phi,k}(Q)}{h_k(Q)}\partial_k\lambda_{\psi,k}(Q)|_{\phi_k}.
\end{equation}
As a net result, we have traded the ``regeneration'' of a four fermion coupling against a modification of a running Yukawa coupling.

Even though no {\em direct} four fermion coupling is now present in the truncation \eqref{eq:trunc_lambda} ($\lambda_{\psi,k}\equiv 0$), the effective four fermion coupling can always be reconstructed from the partially bosonised effective action. For this purpose one solves the field equation for $\phi_k$ as a functional of $\psi$ and reinserts the solution $\phi_k^{(0)}[\psi]$ into the effective action. For the truncation \eqref{eq:trunc_lambda} this yields
\begin{equation}
  \label{eq:sol_feq}
  \begin{split}
    \phi_k^{(0)}(Q) &= h_k(Q) P_{\phi,k}^{-1}(Q)\tilde\phi(k), \\
    \lambda_{\psi,k}^\text{eff}(Q) &= -\tfrac 12 h_k^2(Q) P_{\phi,k}^{-1}(Q).
  \end{split}
\end{equation}
Adding the various bosonic channels which have been taken into account in the bosonisation procedure a suitable renormalised local quartic fermion coupling $U_k$ can be defined as a linear combination of $\lambda_{\psi,k}^\text{eff}(Q=0)$. Once the flow of $h_k$ and $P_{\phi,k}$ is computed one can compare the flow of $U_k$ with the flow computed directly in a purely fermionic formulation \cite{Zan97,Hal00,Hon01}. We have checked that with our rebosonisation prescription the ``fermionic'' and ``bosonic'' flow of $U_k$ agree in lowest nontrivial order (e.g. $\partial_k U_k=cU_k^2$). We emphasize that this ``one loop accuracy'' of the flow of $U_k$ would {\em not} hold had we omitted the ``rebosonisation correction'' \eqref{eq:rebos_corr}.

Beyond one loop order the ``fermionic'' and ``bosonic'' flow of $U_k$ shows differences that depend on the precise truncation. (Without truncations the two versions of the flow have, of course, to coincide precisely.) They reflect the emphasis on different aspects of the model. The bosonic flow investigated here takes also into account the effect of the bosonic fluctuations. They contribute to the generation of quartic bosonic couplings which would correspond to eight--fermion--vertices in a purely fermionic language. These quartic bosonic interactions are crucial for our ability to explore spontaneous symmetry breaking by the renormalisation group flow. Indeed, they guarantee that the bosonic effective potential remains bounded from below even if the bosonic propagator $P_{\phi,k}(Q)$ turns negative for some range of $Q$. From equation \eqref{eq:sol_feq} we learn that the effective four fermion coupling $U_k$ will diverge precisely when $P_{\phi,k}$ vanishes. This divergence limits the the effective four fermion coupling of the purely fermionic flow to the symmetric phase. Going beyond, and in particular exploring the ordered phase, would require to take at least eight-fermion vertices into account, rendering a purely fermionic description complicated. These difficulties are overcome by the bosonic flow equations which can easily follow the flow of a four-boson-vertex.

On the other hand, we observe that the four fermion vertex multiplied by $\lambda_{\psi,k}^\text{eff}$ (eq. \eqref{eq:sol_feq}) shows a particular structure in momentum space: from the combination $\sum_Q \lambda_{\psi,k}(Q)\tilde\phi(Q)\tilde\phi(-Q)$ the momentum of a pair of two fermions sum up to a common ``pair momentum'' $Q$. This ``factorised structure'' does not exhibit the most general momentum dependence of the effective four fermion vertex permitted by the symmetries. It is therefore not possible to rebosonise the most general momentum dependent four fermion interaction generated by the flow. As long as one uses only local bosonic fields only the ``dominant part'' of the four fermion interaction -- precisely the one that can be described by boson exchange in several channels -- can be made to vanish for all $k$ with the help of field redefinitions. In our approximation, we simply omit the ``subdominant part'' that does not take a factorised form. This contrasts with the work on the fermionic renormalisation flow \cite{Zan97,Hal00,Hon01,Gro01} where, in principle, the full momentum dependence of the four fermion interaction is taken into account. (Of course, one could combine the advantages of both approaches by extending the truncation, keeping explicitly the influence of the ``subdominant'' four fermion interactions.) In short, whereas the fermionic formulation typically allows for a higher resolution of the momentum dependence of the four fermion vertex, the partially bosonised version permits the inclusion of higher vertices and the investigation of spontaneous symmetry breaking.

\section{Renormalisation flow of the Hubbard model}
\begin{fmffile}{zmf_graph_running}
\label{sec:trunc_antiferr}

We now want to apply the renormalisation group formalism summarized in the preceeding sections to the Hubbard model. We will concentrate on the region close to half filling and low temperatures where the system is dominated by the antiferromagnetic spin density. In this section we will therefore leave aside all bosons apart from the spin density $\vec m(X)$. There is then no ambiguity how the parameter $U_m$ is related to the original fermionic coupling $U$, namely $U_m=U/3$. We extend this truncation by including other bosonic degrees of freedom in section \ref{sec:remove_ambig}.
We recall this in this setting the mean field approximation is not consistent with Hartree-Fock or lowest order perturbation theory
$(U_m\neq U/2)$. The consistency will be established only by the inclusion of the bosonic fluctuation effects.

\vspace{0.5cm}
\noindent
{\bf Truncation}

Let us now try to define a suitable truncation for the effective action. The initial condition (at very large $k$) for the flow equation \eqref{eq:flowAEA} is the classical action. In the course of the flow towards lower scales the effective average action will in general pick up all possible couplings that are compatible with the symmetries of the theory. In order to make progress we have to truncate this set of infinitely many couplings. We will make an ansatz containing a fermionic kinetic term $\Gamma_{\psi,k}$, a term containing a Yukawa like interaction between fermions and bosons $\Gamma_{Y,k}$ and a bosonic action to be specified below. (A term containing a four fermion interaction is to be rebosonised as sketched in the previous section.) As we are mainly interested in antiferromagnetic behaviour we define the boson
\begin{equation}
  \vec a(Q) = \vec m(Q+\Pi),
\end{equation}
whose zero momentum mode $\vec a(0)$ corresponds to a homogeneous antiferromagnetic spin density.

For the fermionic kinetic term we adopt the classical part unchanged
\begin{eqnarray}
  \label{eq:trunk_ferm}
&&\Gamma_{\psi,k}[\psi,\psi^*] = \sum_Q \psi^\dagger(Q) P_F(Q) \psi(Q),\\
&&P_F(Q) = i\omega_F+\epsilon-\mu, \quad \epsilon(\Bs q) = -2t(\cos q_x+\cos q_y),\nonumber
\end{eqnarray}
where we restrict ourselves to nearest neighbour hopping. In contrast, the Yukawa coupling $\bar h_{a,k}$ is taken to be scale dependent
\begin{eqnarray}\label{eq:trunk_yuk_I}
\Gamma_{Y,k}[\psi,\psi^*,\vec a] =-\bar h_{a,k}\sum_{KQQ'}&&\vec a(K)\psi^*(Q)\vec\sigma\psi(Q')\\
&&\times\delta(K-Q+Q'+\Pi).\nonumber
\end{eqnarray}
For the purely bosonic part we investigate a kinetic term plus a local effective potential
\begin{equation}
  \label{eq:trunk_bos_I}
  \Gamma_{a,k}[\vec a] = \tfrac 12\sum_Q\vec a(-Q) P_a(Q)\vec a(Q) + \sum_X U[\vec a(X)].
\end{equation}
Due to $SU(2)$ symmetry the potential can only depend on the rotation invariant combination $\alpha(X) = \frac 12 \vec a(X)\vec a(X)$ or, in momentum space
\begin{equation}
  \alpha(K,K') = \tfrac 12 \vec a(K) \vec a(K').
\end{equation}
Furthermore we will make an expansion in powers of the field $\vec a$ up to a quartic interaction. We take a different expansion point in the symmetric regime (SYM) and in the regime with spontaneous symmetry breaking (SSB) as it is preferable to always expand around the minimum of the potential

\begin{eqnarray}\label{eq:trunc_U}
\text{SYM} : \sum_X U[\vec a] &=& \sum_K \bar m_a^2\alpha(-K,K)+{}\\
&&+\tfrac 12\sum_{K_1\ldots K_4} \bar\lambda_a \delta(K_1+K_2+K_3+K_4)\nonumber\\
&& \hspace{0.8cm}\times\alpha(K_1,K_2)\alpha(K_3,K_4),\nonumber\\
\text{SSB} : \sum_X U[\vec a] &=& \tfrac 12 \sum_{K_1\ldots K_4} \bar\lambda_a \delta(K_1+K_2+K_3+K_4)\nonumber\\
&&\hspace{0.8cm}\times(\alpha(K_1,K_2)-\alpha_0\delta(K_1)\delta(K_2))\nonumber\\
&&\hspace{0.8cm}\times(\alpha(K_3,K_4)-\alpha_0\delta(K_3)\delta(K_4))\nonumber
\end{eqnarray}
The parameters in the potential are the
scale dependent mass $\bar m_a(k)$ or the minimum $\alpha_0(k)$, as well as the coupling $\bar\lambda_a(k)$. For large $k$ we start in the SYM--regime and switch to the flow equations for the SSB--regime at the scale $k_\text{SSB}$ defined by $\bar m_a^2(k_\text{SSB})=0$.

The bosonic propagator on the classical level is simply a mass term in our case, i.e. $P_a=0$, $\bar\lambda_a=0$. We let us guide by the one loop results of the preceding section for the momentum dependence and take
\begin{equation}
  \label{eq:trunk_bosprop}
  P_a(Q) = Z_a \hat Q^2 = Z_a (\omega_B^2 + t^2 [\Bs q]^2),
\end{equation}
where $Z_a$ is a scale dependent wave function renormalisation (with the dimension of $\text{mass}^{-1}$) and the function $[\Bs q]^2$ is defined as $[\Bs q]^2 = q_x^2 + q_y^2$ for $q_i\in [-\pi,\pi]$ and continued periodically otherwise.

The ansatz $[\Bs q]^2$ for the spatial part is to mimic the momentum dependence of figure \ref{fig:oneloop_prop_m_k}. It is not very accurate for $q_i^2$ in the vicinity of $\pi^2$. For $k\ll\Lambda$ this region gives only a small contribution, however, whereas for $k$ near $\Lambda$ the whole kinetic term is small as compared to the mass term. (The classical approximation is $Z_a=0$.) The dependence on the Matsubara frequency is more difficult. A look at figure \ref{fig:oneloop_prop_m_mats} shows that the bosonic propagator is changed only for the smallest frequencies, while the higher frequency modes retain the classical mass term (without sizeable kinetic term). This could be reflected by the $\omega_B$--dependent part $X_a(\omega_B)$ in $P_a$ obeying $X_a(0)=0$, $X_a \approx m_\text{cl}^2 - \bar m_a^2(k)$ for large $\omega_B$. Lowering the scale, the $\omega_B=0$ mode will then dominate the propagation more and more. Our approximation $X_a(\omega_B)=Z_a\omega_B^2$ is chosen for the sake of simplicity: it has the correct qualitative property of suppressing the effect of the high frequency modes.

One would suspect a similar behaviour for the quartic boson coupling $\bar\lambda_a$ which is generated by the fermion loop: the low frequency modes are supposedly changed most, implying that $\bar\lambda_a$ depends substantially on the frequency $\omega_B$. (For example we could replace $\bar\lambda_a\to\bar\lambda \prod_{i=1}^4f_\lambda(\omega_{B}^i)$ with $f_\lambda(0)=1$ while $f_\lambda=0$ for large $\omega_B^i$, where $i$ labels the external legs of the diagram contributing to the flow of $\bar\lambda$.) We neglect this effect here since the contribution of modes with large $\omega_B$ is suppressed anyhow by the term $\sim Z_a\omega_B^2$ in the inverse propagator.

In addition to the truncation we still have to specify the regulator functions for the renormalisation group equations.

\vspace{0.3cm}
\noindent
{\bf Fermionic regulator}

The fermionic cutoff function is inspired by the fact that at nonvanishing temperature the propagator $P_F(Q) = i\omega_F+\epsilon-\mu$ has no zero--modes. This means that the temperature itself acts as a regulator. We therefore choose
\begin{equation}
  R_k^\psi(Q) = i\omega_F(\tfrac{T_k}{T}-1) = 2\pi i(n_F + \tfrac 12)(T_k-T),
\end{equation}
which has the effect of replacing the temperature $T$ by some function $T_k$ in the fermionic propagator. We specify this function to be ($p\geq 2$)
\begin{gather}
  T_k^p = T^p + k^p,\\
  \partial_kT_k=\left(\frac{k}{T_k}\right)^{p-1}\to
  \begin{cases}
    1 & \text{if } k \gg T \\
    (k/T)^{p-1} & \text{if } k \ll T
  \end{cases},
\end{gather}
such that for large $k$ the flow mimics a change in temperature whereas for $k\ll T$ the modification of the fermion propagator becomes ineffective. For fermionic modes for which the distance of the momentum to the Fermi surface is smaller than $\pi T$ no additional regularisation is needed anyhow.

In short, we want a cutoff that is ineffective for $k\ll T$ and allows us to approach the Fermi surface continuously for $k>T$. We will see in the next section that this requires $p>2$ and we will choose $p=4$.

\vspace{0.3cm}
\noindent
{\bf Bosonic regulator}

For $k\ll T$ possible infrared problems may arise from the fluctuations of long range bosonic modes. This is particularly relevant near a second order phase transition where the bosonic correlation length diverges. Such potential problems are avoided by the bosonic regulator.
We take a ``linear cutoff'' \cite{Lit}
\begin{equation}
  \label{eq:bos_regulator}
  R_k^a(Q) = Z_a(k^2-\hat Q^2) \Theta(k^2-\hat Q^2),
\end{equation}
where $\hat Q^2$ is defined in eq. \eqref{eq:trunk_bosprop}. This leads to a full propagator of the form
\begin{eqnarray}\label{eq:full_bos_prop}
P^a_k(Q)&=&P_a(Q)+R_k^a(Q) \\
&=& Z_a \big(\hat Q^2\;\Theta(\hat Q^2-k^2)+k^2\;\Theta(k^2-\hat Q^2)\big) = Z_a Q_k^2\nonumber
\end{eqnarray}
where
\begin{eqnarray}\label{eq:full_bos_prop1}
Q_k^2 &=& \left\{\begin{array}{lll}
\hat Q^2  &\text{if } &\hat Q^2>k^2 \\
k^2 & \text{if } &\hat Q^2<k^2.
\end{array}\right.
\end{eqnarray}
The regulator function \eqref{eq:bos_regulator} thus only hampers the propagation of
modes with small momenta and Matsubara frequencies where $\hat Q^2<k^2$. Therefore, by
lowering $k$, we effectively average over larger and larger regions in position space.
We may therefore relate properties of the effective average action $\Gamma_k$ at a given
scale $k$ to properties of size $1/k$ in position space. Since the fermionic modes
are effectively integrated out for $k\ll T$ we can associate in this range $k^{-1}$ with
a typical (linear) size of the experimental probe. Indeed, fluctuations with size larger
than the experimental probe should not be included in a ``physical'' definition of the
relevant effective action. For practical purposes one is therefore
interested in macroscopically large but not infinite values of $k^{-1}$.

\vspace{0.3cm}
\noindent
{\bf Initial values}

For $k\to\infty$ both the fermionic and bosonic propagators vanish. Therefore the
loop contributions vanish and $\Gamma_{k\to\infty}$ indeed equals the classical action.
The initial conditions for the parameters occurring in the truncation can therefore be
read off from the classical action \eqref{eq:action}. For very large $k$ one starts in
the SYM regime with
\begin{equation}
  \bar m_a^2 = U_m, \quad \bar\lambda_a = 0, \quad \bar h_a = U_m, \quad Z_a = 0
\end{equation}

\section{Flow equations at half filling}
\label{sec:halffill}

In this and the following section we study the Hubbard model at half filling, i.e. for vanishing chemical potential $\mu=0$. As a test of the reliability of our approach we extend the truncation in sect. \ref{sec:remove_ambig} and check the influence of the Fierz ambiguity.

\vspace{0.3cm}
\noindent
{\bf Bosonic potential}

The flow equation for the effective potential $U(\alpha)$ obtains by evaluating the ERGE \eqref{eq:flowAEA} for a homogeneous antiferromagnetic order parameter $\vec a(Q) = \vec a \delta(Q)$, $\alpha=\frac 12 \vec a^2$. The contribution of the fermionic fluctuations can be found by replacing in the mean field theory result \eqref{eq:mfUmats} $T\to T_k$ in the fermionic propagator and applying the $\tilde\partial_k$--derivative. The bosonic contribution is the same as for the $O(3)$ linear $\sigma$--model \cite{Wet91,Ber02}:
\begin{align}
  \label{eq:EffPotA}
  \partial_k U & (\alpha) = \partial_k U^B(\alpha) + \partial_k U^F(\alpha) \nonumber \\
  &= \tfrac 12 \sum_{Q,i} \tilde\partial_k \ln[P_a(Q)+\hat M_i^2(\alpha)+R^a_{k}(Q)] \\
  &\qquad - 2T\int_{-\pi}^{\pi} \frac{d^2q}{(2\pi)^2} \tilde\partial_k\ln\cosh y(\alpha). \nonumber
\end{align}
Here the boson mass terms $\hat M_i^2$ are different for the ``radial mode'' (in the direction of $\vec a$) and the ``Goldstone modes'' (perpendicular to $\vec a$),
%
%
%
\begin{equation}
  \label{eq:bos_mass}
 \begin{split}
  &\hat M_{1,2,3}^2(\alpha) ={}\\
    & \qquad =\left\{
    \begin{array}{ll}
      (\bar m_a^2+3\bar\lambda_a\alpha,\bar m_a^2+\bar\lambda_a\alpha,\bar m_a^2+\bar\lambda_a\alpha)&\text{SYM}\\
      (\bar\lambda_a(3\alpha-\alpha_0), \bar\lambda_a(\alpha-\alpha_0), \bar\lambda_a(\alpha-\alpha_0) )&\text{SSB}
    \end{array}
  \right.
 \end{split}
\end{equation}
In the SSB regime we recognize the two massless Goldstone bosons for $\alpha=\alpha_0$. The squared fermion mass (mass gap) $2\bar h_a^2\alpha$ enters the function
\begin{equation}
  y(\alpha) = \frac{1}{2T_k}\sqrt{\epsilon^2(\Bs q) + 2\bar h_a^2 \alpha}.
\end{equation}

The flow of the mass term and quartic coupling follow from simple differentiation of eq.
(\ref{eq:EffPotA})
\begin{equation}
  \label{eq:def_param_U}
  \begin{aligned}
    \text{SYM: }&& \partial_k \bar m_a^2 &= \tfrac{\partial}{\partial\alpha}(\partial_k U(\alpha))|_{\alpha=0},\\
    &&\partial_k \bar \lambda_a &= \tfrac{\partial^2}{\partial\alpha^2}(\partial_k U(\alpha))|_{\alpha=0},\\
    \text{SSB: }&& \partial_k \alpha_0 &= -\tfrac{1}{\bar\lambda_a} \tfrac{\partial}{\partial\alpha}(\partial_k U(\alpha))|_{\alpha=\alpha_0},\quad\\
    &&\partial_k \bar \lambda_a &= \tfrac{\partial^2}{\partial\alpha^2}(\partial_k U(\alpha))|_{\alpha=\alpha_0}.
  \end{aligned}
\end{equation}
Here the flow of the minimum is inferred from the fact that $U'(\alpha_0)=0$ and hence
$\tfrac{d}{dk}U'(\alpha_0)=\partial_kU'(\alpha_0)+U''(\alpha_0)\partial_k\alpha_0=0$.

With the aid of \eqref{eq:def_param_U} we may now derive the flow equations for the parameters in the effective potential. However, first we introduce rescaled and renormalised quantities
\begin{equation}
  u=\frac{U t^2}{T k^2}, \quad \tilde\alpha = \frac{Z_a t^2 \alpha}{T}
\end{equation}
such that
\begin{equation}
  \label{eq:resc_ren_quant}
  \begin{gathered}
    m_a^2 = \frac{\bar m_a^2}{Z_a k^2} = \frac{\partial u}{\partial\tilde\alpha}, \quad
    \kappa_a = \frac{Z_a t^2}{T}\alpha_0,\\
    \lambda_a = \frac{T}{Z_a^2 t^2 k^2}\bar\lambda_a = \frac{\partial^2 u}{\partial\tilde\alpha^2}, \quad
    h_a^2 = \frac{T}{Z_a t^4}\bar h_a^2
    \end{gathered}
\end{equation}
are all dimensionless.

\begin{figure}[htb!]
\centering
\psfrag{-t}{\hspace{-1.5cm}$-\ln k/t$}
\includegraphics[scale=.65]{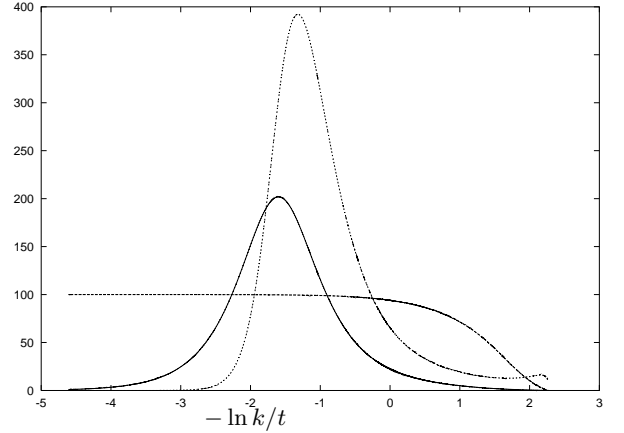}
\caption{Flow of the renormalised and unrenormalised mass $10^{-1}m_a^2$ (solid), $10^{2}\bar m_a^2/t$ (long dashes) and the quartic bosonic coupling $\lambda_a$ (short dashes) in the symmetric regime (SYM) at half filling. We have chosen $U/t=3$ and $T/t=0.15$.}
\label{fig:ayuk_sym}
\end{figure}

\noindent
The flow equation for $u$ can be expressed in terms of ``threshold functions'' $\bar l_0^2(w)$ which express the decoupling of particles with mass much larger than $k$ (large $w$)
\begin{eqnarray}\label{eq:flow_u}
k\partial_k u|_{\tilde\alpha} &=& -2u+\eta_a\tilde\alpha u' +
\frac{t^2}{Tk^2} k\partial_k U|_\alpha\nonumber\\
=&-&2u+\eta_a\tilde\alpha u' + \frac{1}{4\pi}\sum_i
\bar l_0^2(w_i,\eta_a,T)\nonumber\\
&+& s_0^F(\tilde m_F^2, T).
\end{eqnarray}
This uses $\partial_k u|_{\tilde\alpha} = \partial_k u|_\alpha + \eta_a\tilde\alpha u'$, $u' = \partial u/\partial \tilde\alpha$ and $\eta_a$ is the anomalous dimension
\begin{equation}
  \eta_a = -k\partial_k \ln Z_a.
\end{equation}
The dimensionless threshold functions ($\tilde q = tq$)

%
\begin{equation}
  \label{eq:threshold}
  \begin{split}
    \bar l_0^2(w,\eta_a,T) &= \frac{2\pi}{k}\sum_n \int_{-t\pi}^{t\pi}\frac{d^2\tilde q}{(2\pi)^2} \tilde\partial_k \ln(P_k^a+\hat M^2),\\
    s_0^F(\tilde m_F^2, T) &= \left(\frac{k}{T_k}\right)^{p-2}\left(\frac{t}{T_k}\right)^3\\
                           &\hspace{0.5cm}\times\int_{-\pi}^{\pi}\frac{d^2 q}{(2\pi)^2} \tilde y \tanh\left(\frac{t}{2T_k}\tilde y\right),\\
    \tilde y &= \big(4(\cos q_1 + \cos q_2)^2+\tilde m_F^2\big)^{1/2}
  \end{split}
\end{equation}
decay rapidly for large values of the arguments $w=\hat M^2/(Z_ak^2)$, $\tilde m_F^2=2h_a^2\tilde\alpha$. Their explicit form is given in the appendix.
The characteristic scale dependence of the mass term and quartic
coupling is shown in Fig. \ref{fig:ayuk_sym}. The running for $k\gg t$ has
not much physical meaning and rather sets the stage for the correct translation
between the classical action and $\Gamma_{k\approx t}$. The physical fluctuation
effects become important for $k<t$ as seen in the running of $\bar{m}^2_a$.

\begin{figure}[!htb]
\label{eq:feyn_mass}
\includegraphics[width=6cm]{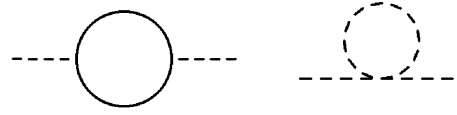}
\caption{Diagrams for the running of the scalar two point function.}
\label{Fig.7}
\end{figure}

It is instructive to consider the long wavelength limit $k^2\ll T^2$. For $p>2$ the fermionic contribution $s_0^F$ vanishes $\sim k^{p-2}$. As argued in the previous section the fermionic infrared cutoff should be ineffective for $k\ll T$. This means that the change in $T_k$ becomes irrelevant. In turn, this requires sufficiently large $p$ and we choose $p=4$. On the other hand the bosonic threshold function approaches for $k\ll T$ the one for the two dimensional linear $O(3)$ $\sigma$--model \cite{Ber02,Lit,Hoe02}
\begin{equation}
\bar l_0^2(w,\eta_a,T) \to l_0^2(w,\eta_a) = \frac{1-\eta_a/4}{1+w}.
\end{equation}
In the long wavelength limit we therefore recover the well studied flow equations for the scalar $O(3)$ model in two dimensions \cite{Ger01}. The flow equations for $m_a^2$, $\lambda_a$ and $\kappa$ involve the $\alpha$--derivatives of $l_0^2$ and $s_0^F$ and can be found in the appendix. We note that all these equations have a simple diagrammatic representation. For example, the mass contributions in the symmetric regime are shown in Fig. \ref{Fig.7}.

\vspace{0.3cm}
\noindent
{\bf Anomalous dimension}

The anomalous dimension $\eta_a = -k\partial_k\ln Z_a$ is a measure for the change of the wave function renormalisation $Z_a$ with scale. Therefore we can extract it from the momentum dependence of the bosonic two point function.
\begin{equation}\label{116}
  \begin{split}
    \Gamma_{B,\text{kin}} &= \sum_K P_a(K) \alpha(-K,K) = Z_a \sum_K {\hat K}^2 \alpha(-K,K).
  \end{split}
\end{equation}
As we are mainly interested in the spatial momentum dependence we define
\begin{gather}\label{117}
  Z_a = \frac{1}{t^2} \frac{\partial}{\partial (l^2)}\bigg|_{l=0}\bigg\{
  \frac{\delta}{\delta \alpha(-K,K)}\Gamma_k
  \bigg\}\bigg|_{\psi,\psi^*=0,\alpha =\alpha_0},\\
  K=(\omega_B=0,\Bs k = l\Bs e_1), \nonumber
\end{gather}
where $\alpha_0$ is the minimum of the effective potential. We observe that $Z_a$ is
closely related to the momentum dependence of the inverse propagator shown in
figs. 3, 4. The definition (\ref{116}), (\ref{117}) corresponds to an approximation
near the minimum, where the constant part (at the minimum) is substracted.

\begin{figure}[htb!]
\centering
\psfrag{-t}{\hspace{-1.5cm}$-\ln k/t$}
\includegraphics[scale=.65]{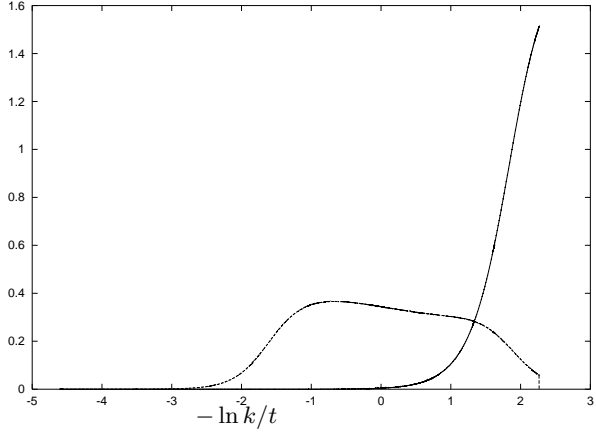}
\caption{Flow of the wave function renormalisation $Z_a$ (solid) and the anomalous dimension $10^{-1}\eta_a$ (dashes) in the symmetric regime (SYM) at half filling. We have chosen $U/t=3$ and $T/t=0.15$.}
\label{fig:Za_eta_sym}
\end{figure}

In the symmetric phase (SYM) the bosonic propagator is affected by the two diagrams in Fig. 7. However, the bosonic loop is independent of the external momenta and therefore does not contribute to the anomalous dimension. The fermionic loop is well known from our loop calculation by the replacements $T\to T_k$, $U_m\to\bar h_a$ for $P_a(K)=P_m(K+\Pi)$ in eq.\eqref{eq:bos_prop_korr}, and performing the $\tilde\partial_k$--derivative. We obtain
%
\begin{eqnarray}\label{eq:eta_sym}
&\eta_a^{(F)} = h_a^2 t (k\partial_kT_k)
\Big[\frac{\partial}{\partial T_k}\frac{\partial}{\partial(l^2)}\\
&\hspace{1.5cm}\Big\{\frac{1}{T_k}\int_{-\pi}^{\pi} \frac{d^2q}{(2\pi)^2}
\frac{\tanh\tfrac{\epsilon(\Bs q)}{2T_k}+\tanh\tfrac{\epsilon(\Bs q+l\hat e_1)}{2T_k}}
{(\epsilon(\Bs q)+\epsilon(\Bs q+l\hat e_1))/t}
\Big\}\Big]_{l=0}.\nonumber
\end{eqnarray}
A characteristic evolution of $Z_a$ and $\eta_a$ is plotted in Fig. \ref{fig:Za_eta_sym}.

In the $SSB$-regime we also get a contribution from the bosonic sector. The contribution comes from a bosonic loop with four external legs, where two external legs are connected to the condensate.  As we will see later, the system enters the broken phase at small values of the cutoff parameter $k$. At these values only the lowest Matsubara frequency ($\omega_B=0$) contributes in the bosonic propagator (c.f. \eqref{eq:full_bos_prop}). The bosonic contribution to $\eta_a$ is then the same as the corresponding one for a simple $O(3)$ model in two dimensions. Here the anomalous dimension has been calculated to be \cite{Wet91}
\begin{equation}
  \eta_a^{(B)} = \frac{\lambda_a^2\kappa_a}{\pi} m_{2,2}^2(2\lambda_a\kappa_a,0)
  = \frac{\lambda_a^2\kappa_a}{\pi(1+2\lambda_a\kappa_a)^2}, \label{eq:eta_SSB}
\end{equation}
where the function $m_{2,2}^2(w_1,w_2)$ contains the momentum dependence of the loop integral and depends on the regulator $R_k^a$. For our choice of regulator it has been computed explicitely \cite{Hoe02} as shown in \eqref{eq:eta_SSB}. In the $SSB$-regime the contribution from the fermionic loop is much smaller than the bosonic contribution and we will neglect it.


\vspace{0.3cm}
\noindent
{\bf Running Yukawa coupling}

\begin{figure}[h]
\[
\begin{gathered}
 \includegraphics[width=6cm]{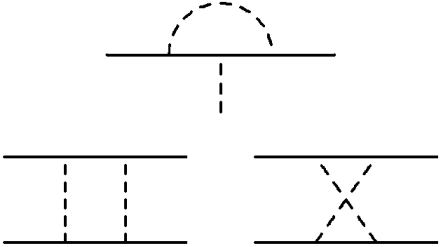}
  \label{feyn:coupling}
\end{gathered}
\]
\caption{Diagrams for the running Yukawa coupling}
\label{Fig.9}
\end{figure}

In the symmetric phase the running of the Yukawa coupling is generated by the diagrams in Fig. \ref{Fig.9}.

The first diagram is the direct contribution, while the last two contribute by rebosonisation as prescribed in sect. \ref{sec:rebos}. More details on the extraction of the rebosonised flow equations can be found in appendix \ref{app:rebos}.

In the symmetric phase we get for the running of $h_a$
\begin{equation}
  \label{eq:floweq_ha}
  k\partial_k h_a^2 = h_a^2 \eta_a + \beta_{h^2_a}^{(d)} + \beta_{h^2_a}^{(rb)}
\end{equation}
where the ``direct'' and ``rebosonised''
contributions to the beta functions read $(\hat{\omega}_F=(2n+1)\pi T_k)$
%
\begin{equation}
  \label{eq:betafun_ha}
  \begin{split}
    \beta_{h^2_a}^{(d)}
    =& - 2 h_a^4 \frac{t^2}{k^2 T} k\tilde\partial_k \sum_Q\big\{ p_a^{-1}(Q) p_F^{-1}(Q)p_F^{-1}(Q+\Pi) \big\} \\
    =& - 2 h_a^4 \frac{t^4}{k^2 T} \sum_Q\bigg\{\frac{k\partial_k \hat{\omega}_F^2}{(\hat{\omega}_F^2+\epsilon^2(\Bs q))^2}\frac{1}{Q_k^2/k^2+m_a^2}+{}\\
    &+ \frac{1}{\hat{\omega}_F^2+\epsilon^2(\Bs q)}\frac{(2-\eta_a(1-{\hat Q}^2/k^2))\Theta(k^2-{\hat Q}^2)}{(1+m_a^2)^2}
    \bigg\}
    \\
    \beta_{h^2_a}^{(rb)}
    =& 2m_a^2 h_a^4 \frac{t^2}{k^2 T} k\tilde\partial_k\sum_Q\big\{p_a^{-1}(Q)p_a^{-1}(\Pi-Q)p_F^{-1}(-Q)\\
    &\hspace{3.0cm} \times [p_F^{-1}(\Pi-Q)-p_F^{-1}(Q)]\big\}\\
    =& 4m_a^2 h_a^4 \frac{t^4}{k^2 T} \sum_Q\bigg\{\frac{k\partial_k \hat{\omega}_F^2}{(\hat{\omega}_F^2+\epsilon^2(\Bs q))^2}\frac{1}{Q_k^2/k^2+m_a^2}+{}\\
    &+ 2\frac{1}{\hat{\omega}_F^2+\epsilon^2(\Bs q)}\frac{(2-\eta_a(1-{\hat Q}^2/k^2))\Theta(k^2-{\hat Q}^2)}{(1+m_a^2)^2}
    \bigg\}\\
    &\hspace{3.5cm} \times \frac{1}{(Q+\Pi)_k^2/k^2+m_a^2}
  \end{split}
\end{equation}
In eq. (\ref{eq:betafun_ha}) we use $Q_k^2 = {\hat Q}^2\Theta({\hat Q}^2-k^2)+k^2\Theta(k^2-{\hat Q}^2)$, $k\partial_k\hat{\omega}_F = (2n+1)\pi k^p T_k^{1-p}$ and
\begin{equation}
  p_F(Q) = (i\hat{\omega}_F+\epsilon({\bf q}))/t,\quad
  p_a(Q) = Q_k^2/k^2 + m_a^2.
\end{equation}
We plot the flow of the renormalised and unrenormalised Yukawa coupling in the symmetric regime in fig. 10.

\begin{figure}[h!tb!]
\centering
\psfrag{-t}{\hspace{-1.5cm}$-\ln k/t$}
\includegraphics[scale=.65]{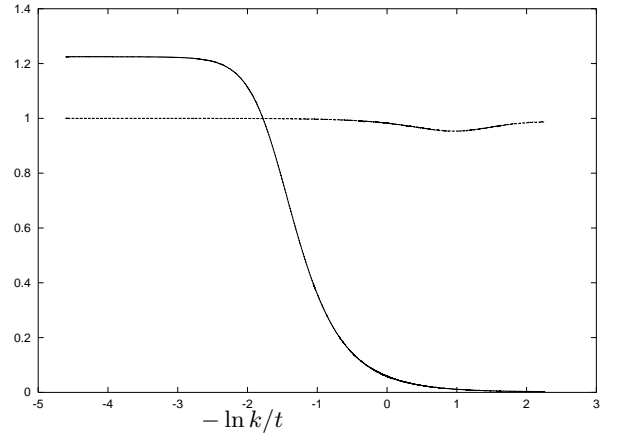}
\caption{Flow of the renormalised $10^{-2}h_a$ (solid) and unrenormalised $\bar h_a$ (dashes) Yukawa couplings in the symmetric regime (SYM) at half filling. We have chosen $U/t=3$ and $T/t=0.15$.}
\label{fig:ayuk_sym_h}
\end{figure}

For the specific parameters used for this plot we observe very little running of $\bar h_a$. The charge will be more pronounced for other choices of the parameters, as discussed in sect. XV.
In the regime with spontaneously broken $SU(2)$--symmetry (SSB) the change in the Yukawa coupling is negligible as we have checked numerically. We neglect the running of the Yukawa--coupling $\bar h_a(k)$ in the SSB regime. For $k<k_\text{SSB}$ we therefore keep the unrenormalised Yukawa coupling fixed $\bar h_a(k<k_\text{SSB})=\bar h_a(k_\text{SSB})$.

\section{Spontaneous symmetry breaking}
\label{sec:spont_symm_break}

\begin{figure}[tb!]
\centering
\psfrag{-t}{\hspace{-1.5cm}$-\ln k/t$}
\includegraphics[scale=.65]{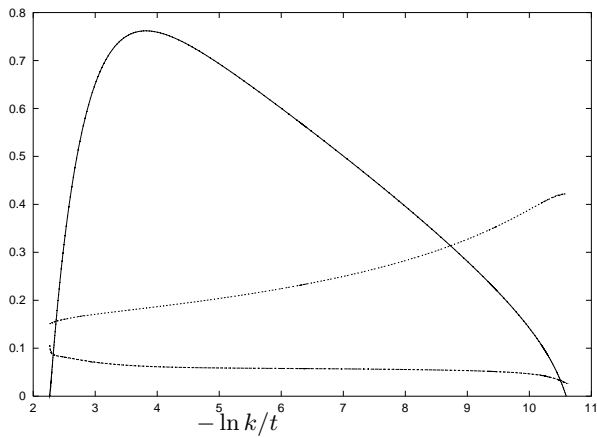}
\caption{Flow of the renormalised minimum of the potential $\kappa_a$ (solid), the quartic bosonic coupling $10^{-2}\lambda_a$ (long dashes) and the wave function renormalisation $10^{-1}Z_a$ (short dashes) in the broken (SSB) regime at half filling. We have chosen $U/t=3$ and $T/t=0.15$.}
\label{fig:ayuk_ssb}
\end{figure}
\begin{figure}[tb!]
\centering
\psfrag{-t}{\hspace{-1.5cm}$-\ln k/t$}
\includegraphics[scale=.65]{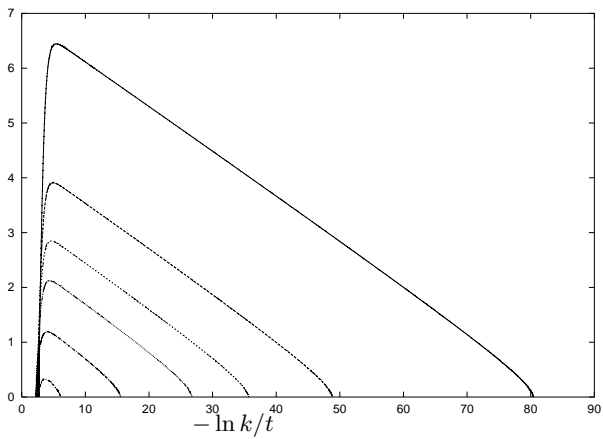}
\caption{Flow of the renormalised minimum of the potential $\kappa_a$ in the broken
(SSB) regime at half filling for different temperatures. We have chosen $U/t=3$.
The temperatures are in the set $T/t=\{0.05, 0.07, 0.085, 0.10, 0.13, 0.17\}$
(from top to bottom). For $T/t=0.05$ the order persists up to
$k\approx 10^{-40} t$ corresponding
to a sample size of $10^{31}$ cm. This clearly demonstrates that the Mermin-Wagner theorem is not
practically applicable.}
\label{fig:ayuk_ssb_multi}
\end{figure}

We now turn to a numerical analysis of the above flow equations. For this we set $U/t=3$ and take a temperature $T=0.15t$. The initial scale $k_0=100t$ is chosen so large that the final results do not depend on it and the one loop results are well produced in the beginning of the flow. The differential equations were integrated by a standard Runge--Kutta like routine \cite{Pre92}. In fig. 6 we plot the flow of the renormalised mass $m_a^2$, the unrenormalised mass $\bar m_a^2$ and the quartic bosonic coupling $\lambda_a$. The values of $k$ correspond to a flow in the symmetric regime (SYM) where the minimum of the effective potential occurs for a vanishing order parameter $\alpha=0$. One observes that the running of $\bar m_a$ sets in only for $k\lesssim t$. Then both $\bar m_a$ and $m_a$ decrease. At $\ln k/t\approx -2.2$ the bosonic mass $m_a^2$ vanishes and for smaller values of $k$ we enter the SSB regime. We denote the corresponding scale by $k_\text{SSB}$. In figure 8 we plot the flow of the wave function renormalisation $Z_a$ and the corresponding anomalous dimension $\eta_a$. This explains qualitatively the flow of the renormalised couplings in fig. 6: for scales above $\ln k/t \approx 2.0$ the running is mainly dominated by the simple scaling due to the respective dimensions of the couplings. In an intermediate range down to $\ln k/t \approx -1$ the large value of $\eta_a$ dominates the flow while for even smaller values of $\ln k/t$ the fermionic part of the flow equations play an ever increasing role.

In figure 11 we explore the SSB regime with $\alpha(k)\neq 0$ for $k<k_\text{SSB}$. First, we observe that the quartic bosonic coupling reaches a fixed point very soon. This is because the positive term $\sim\lambda_a^2$ in equation \eqref{eq:flow_lambda_ssb} just compensates the negative contributions from the fermions and dimension-scaling. Two opposite effects govern the flow of the location of the minimum of the potential, as parameterized by $\kappa_a$. For $k$ only somewhat smaller than $k_\text{SSB}$ the fermionic fluctuations dominate the flow. They lead to increasing values of $\kappa_a$. However, soon the fermionic contribution becomes smaller due to the increasing mass (gap) of the fermions. Then the bosonic loop involving the massless Goldstone bosons dominates. This only results in a slow logarithmic ``running'' of $\kappa_a$, but it finally drives the minimum to zero and thus restores the symmetry. When the fermionic part becomes negligible we effectively deal with a bosonic $O(3)$ model in two dimensions for which the symmetry restoration is a well known feature and has been extensively studied in our functional renormalisation group formalism \cite{Wet91,Ber02,Ger01}.

This picture reconciles the antiferromagnetic order with the Mermin--Wagner theorem \cite{Mer66}, which states that a continuous symmetry cannot be broken at nonvanishing temperature in two dimensions and below. We indeed loose the notion of symmetry breaking if we average over arbitrarily large volumes, i.e. lower the cutoff parameter $k$ to zero. For $k\to 0$ the order parameter $\alpha(k)$ always vanishes,
in accordance with the Mermin-Wagner theorem. This is due to the presence of Goldstone modes in the SSB regime\footnotemark. Nevertheless, we find that for low temperatures the symmetry is restored only when averaging over extremely large samples unaccessible to any real experiment.
\footnotetext{For $O(2)$ models in two dimensions there is another possibility to circumvent the Mermin--Wagner theorem related to a
Kosterlitz and Thouless phase transition \cite{Kos73}. It is speculated that this kind of mechanism may play a role in the superconducting region \cite{Bai00,Cha02}.}

For example, for $T=0.05t$ the restoration of the $O(3)$-symmetry would require the effects
of the Goldstone-boson fluctuations with characteristic length scale (inverse momentum)
of $10^{31}cm$. For any realistic experimental length scale, however,
the order parameter $\alpha(k)$ remains nonzero. This includes macroscopic length scales.
We conclude that it makes perfect sense to speak about long range antiferromagnetic order.

Let us consider a probe with some finite macroscopic size, say $1cm$. Obviously,
the fluctuations with momenta smaller than $(1cm)^{-1}$ are absent in this setting and
we should stop the renormalization flow at some nonzero $k=(1cm)^{-1}\approx 10^{-9}t$. For
all observations with characteristic length scale smaller than $1cm$ (in order to avoid
boundary effects from the particular geometry of the probe) one observes all features
of an effective antiferromagnetic state: For all temperatures smaller than a ``critical
temperature'' $T_c$ there will be an average spin direction $\vec{a}_0$
(``spontaneous symmetry breaking''). Furthermore, the fluctuations orthogonal to
$\vec{a}_0$ will behave as two Goldstone modes with effectively ``infinite'' ($\geqslant$
size of the sample) correlation length for all $T<T_c$. On the other hand, the
antiferromagnetic spin wave fluctuations in the direction of $\vec{a}_0$ (radial mode) will
have a finite correlation length $\xi$ for $T<T_c$. As $T_c$ is approached from below
also the radial correlation length diverges
$\xi(T\rightarrow T_c)\rightarrow''\infty''$, i.e. exceeds the size of the probe.
The different behavior of the Goldstone and
radial modes is a direct consequence of the presence of a direction specified by
$\vec{a}_0$. Another consequence of effective antiferromagnetic order is the
presence of a mass gap for the electrons and holes.

Let us next turn to the notion of a critical temperature $T_c$. We will see that its
precise value depends on the size $k^{-1}$ of the experimental probe. For this purpose
we display in Fig. \ref{fig:ayuk_ssb_multi}
the running of $\kappa_a$ for various temperatures. For a fixed
value of $k>0$ (corresponding to the inverse of the sample size) we always observe a
transition to an ordered phase if $T$ is low enough. This transition temperature is
denoted by $T_c(k)$. In fig. 1 it is directly visible as the temperature for which the
order parameter $\kappa_a$ vanishes.
In fact, fig. \ref{fig:ssb} shows
$\kappa_a(k)$ with $k/t=10^{-9}$ much smaller than all other characteristic
scales of the model.

Inspecting fig. \ref{fig:ayuk_ssb_multi} we may define a characteristic scale oof symmetry
restoration $k_\text{SR}$ by the vanishing of $\kappa_a$, i.e. $(\alpha(k_\text{SR})=0$.
Only ``domains'' with a size of $k_\text{SR}^{-1}$ become so randomly
oriented that no net order prevails on length scales larger than $k_\text{SR}^{-1}$.
For an experimental setting with characteristic scale $k \gg k_\text{SR}$ the
formal ``asymptotic
symmetry restoration'' for $k\rightarrow 0$
is without practical relevance. Thus, even though formally correct,
the Mermin--Wagner theorem fails to be practically applicable in the case of the
Hubbard model at low temperature.
In contrast, for $k\ll k_\text{SR}$ the precise value of $k$ is unimportant and we may
take $k=0$. In fact, this situation corresponds to the ordered phase with a
renormalized mass (or inverse correlation length) $m_R\sim k_\text{SR}$ \cite{Ger01}. The
correlation length $\xi$ acts as an effective physical cutoff such that fluctuations
on length scales larger than $\xi$ are irrelelvant and an additional ``experimental
cutoff'' $k\ll m_R$ has no effect.

\section{Critical behavior}
\label{sec:criticalbehavior}
For $T$ sufficiently large compared to $T_c$ our analysis confirms the statement
of \cite{CHN} that the long distance behavior of the Hubbard model can be described by a
classical nonlinear $\sigma$-model. This holds for $T$ in the vicinity of and below $T_c$.
We can use our findings in order to establish a quantitative description of the
temperature dependence of the correlation length $\xi=m^{-1}_R$. For characteristic
momenta $\ll T$ the fermion fluctuations are cut off by the temperature and only the bosonic
flutuations remain relevant. Within the remaining effective classical $O(3)$ linear
$\sigma$-model the running of $\kappa$ obeys for $\kappa$ sufficiently large
\cite{Ger01}
\begin{equation}\label{CS1}
k\partial_k\kappa=\frac{1}{4\pi}+\frac{1}{16\pi^2\kappa}+0(\kappa^{-2}).
\end{equation}
As a reasonable approximation (cf. fig. \ref{fig:ayuk_ssb}) we may employ a ``linear''
\footnote{The nonlinearities result in a propertionality factor
$\tilde{k}_\text{SR}=k_\text{SR}\tilde{C}_\text{SR}$ and in a slight derivation of $k_m$ from the
location of the maximum in fig. \ref{fig:ayuk_ssb_multi}.} behavior for $k$ between $k_m$ and
$\tilde{k}_\text{SR}$
\begin{equation}\label{CS2}
\kappa(k)=\kappa_m(T)-\frac{1}{4\pi}\ln\frac{k_m(T)}{k}
\end{equation}
where $\kappa_m(T)$ corresponds to the maximum of the curves in fig. \ref{fig:ayuk_ssb}.
With $\xi^{-1}=\tilde{k}_\text{SR}/C_\text{SR}$ and $\kappa(\tilde{k}_\text{SR})=0$ one obtains for
the correlation length
\begin{equation}\label{CS3}
\xi=\frac{C_\text{SR}}{k_m(T)}\exp\big(4\pi\kappa_m(T)\big).
\end{equation}
Using the definition (\ref{eq:resc_ren_quant}) for $\kappa$ this yields the characteristic
behavior \cite{CHN}
\begin{equation}\label{CS4}
\xi=\tilde{C}\exp\left(\frac{\gamma}{T}\right)
\end{equation}
with
\begin{equation}\label{CS5}
\gamma=4\pi\hat{\alpha}_0(T)\hat{Z}_a(T)t^2.
\end{equation}
Here $\hat{\alpha}^{(T)}\hat{Z}_a^{(T)}$ denotes the $T$-dependent maximum of
$(\alpha_0Z_a)(k)$ and both
$\gamma$ and $\tilde{C}=C_\text{SR}/k_m$ depend smoothly on $T$ (without particular
features for $T\rightarrow 0)$.

For $T<T_c$ the correlation length according to eq. (\ref{CS4}) exceeds the size of the
experimental probe and eq. (\ref{CS4}) is no longer applicable. The modifications
of the results of \cite{CHN} for $T<T_c$ therefore concern the temperature region
where the formal correlation length (for $k\rightarrow 0)$ would become extremely
large and not practically meaningful anymore. We can actually use eq. (\ref{CS2})
for an estimate of the critical temperature $T_c(k)$
\begin{equation}\label{CS5a}
T_c(k)=\frac{\gamma(T_c)}{\ln\big(k_m(T_c)/k\big)}.
\end{equation}
The $k$-dependence of $T_c$ is only mild - typically a change in $k$ by a factor of $10$
changes $T_c$ by $10$ percent. For a sample size $k=10^{-9}t\approx(1cm)^{-1}$ we
may use a quantitative formula for $\xi$
\begin{equation}\label{CS6}
\xi t=c(T)\exp\left\{20.7\beta(T)\frac{T_c}{T}\right\}
\end{equation}
with
\begin{eqnarray}\label{CS7}
\beta(T)&=&\frac{\hat{\alpha}_0(T)\hat{Z}_a(T)}
{\hat{\alpha}_0(T_c)\hat{Z}_a(T_c)},\nonumber\\
c(T)&=&C_\text{SR}\frac{k_m(T_c)}{k_m(T)}
\left(\frac{k_m(T_c)}{t}\right)^{\delta(T)},\nonumber\\
\delta(T)&=&\beta(T)\frac{T_c}{T}-1
\end{eqnarray}
and $\beta(T_c)=1,c(T_c)=C_\text{SR}$. Despite the fact that $T_c$ is not a
sharp temperature the behavior for $T$ near $T_c$ shares several features of standard
critical behavior. It is obvious that the correlation length increases very rapidly
as $T$ is lowered from say $1.5T_c$ to $1.1T_c$ - still, it does not diverge for
$T\rightarrow T_c$. On the other hand, for $T<T_c$ the order parameter reaches zero for
$T\rightarrow T_c$ (fixed $k$) according to
\begin{equation}\label{CS8}
\kappa_a(T)=\left(\frac{\gamma(T)T_c}{T}-1\right)\kappa_m(T_c)
+\frac{1}{4\pi}\ln\frac{k_m(T_c)}{k_m(T)}.
\end{equation}
The critical temperature $T_c=0.114t$ can be directly read off from fig. \ref{fig:ssb}.
Furthermore, in the vicinity of $T_c$ the propagator $G(q^2)$ for the antiferromagnetic
spin waves with momenta $q^2\gg\xi^{-2}$ is characterized by an anomalous dimension
$\eta_a=\eta_a(k=\sqrt{q^2},T=T_c)$, namely
\begin{equation}\label{CS9}
G(q^2)=\big(Z_a(k=\sqrt{q^2})q^2\big)^{-1}\sim(q^2)^{-1+\eta_a/2}
\end{equation}
We emphasize that all the critical features discussed in this section extend beyond the
particular approximations of this work and beyond the particular next-neighbour-coupling
Hubbard model. They are universal and hold whenever the long distance behaviour can be
described by an effective two dimensional model for a spin field with $O(3)$ symmetry.

Inspection of fig. \ref{fig:ayuk_ssb_multi} reveals another characteristic temperature,
namely the minimal temperature for which the flow enters the SSB-regime. We will call this
the pseudocritical temperature $T_{pc}$ and infer $T_{pc}=0.18t$ (for $U/t=3)$. For
$T>T_{pc}$ the mass term $m^2_a(k)$ remains positive for all $k$. We note that $T_{pc}$
is substantially above $T_c~(T_{pc}/T_c\approx 1.6)$. In the temperature range between
$T_c$ and $T_{pc}$ we may describe the physics by antiferromagnetic domains with
fluctuating directions. The typical size of domains with local order extends up to the
scale $k^{-1}_\text{SR}$ where $\kappa_a(k_\text{SR})=0$. On length scales larger than
$k^{-1}_\text{SR}$ no long range order remains. It is obvious that the critical behaviour
described by eq. (\ref{CS2}) looses its validity for $T$ above $T_{pc}$ or in the
vicinity of $T_{pc}$. Therefore $T_{pc}$ roughly describes the upper bound of the range of
validity in $T$ for the critical behaviour discussed above.

The positivity of $m^2_a$ for $T>T_{pc}$ implies that the effective four-fermi-interaction
mediated by the exchange of the $\vec{a}$-boson remains finite. We conclude that the
``critical temperature'' determined by the fermionic flow equations
\cite{Zan97, Hal00, Hon01,Gro01} actually corresponds to $T_{pc}$ rather than $T_c$.
Indeed, the approach based on the fermionic flow equation precisely looks for the
``critical temperature'' where the four-fermion-interaction diverges for some
$k$ - or, in our language, where $m^2_a$ reaches zero. The same holds for mean field
theory or the Hartree-Fock approximation. Again, these approaches determine $T_{pc}$
rather than $T_c$. The true critical behaviour discussed in this section is
governed by the renormalisation flow for the effective interactions of composite
Goldstone bosons. It cannot be captured by the Hartree-Fock approximations or the
fermionic flow equations. We emphasize that the influence of the Goldstone boson
fluctuation on the value of the critical temperature is a large effect - we find
$T_c$ almost a factor of two lower \footnote{A substantial lowering of $T_c$ has also
been observed in a $1/d$ expansion \cite{VD}, with $d$ the dimensionality of space.
The precise connection with our result remains unclear, however, since the $1/d$ expansion
at most partially captures the Goldstone boson physics.} than for the mean field theory.

\section{Renormalisation flow for nonzero doping}
\label{sec:flownonzero}

In this section we want to compute parts of the phase diagram for nonzero doping. For this purpose we study the flow equation for nonvanishing chemical potential in the symmetric phase (i.e. $\mu\neq 0$, $\alpha = 0$). The bosonic part of the flow of the effective potential is not altered, while the fermionic part gets modified. Similar to eq. \eqref{eq:flow_u} we obtain
\begin{equation}
  \label{eq:flow_u_mu}
  \begin{aligned}
    k\partial_k u|_{\tilde\alpha} =& -2u+\eta\tilde\alpha u' \\
    &+ \frac{1}{4\pi}\sum_i \bar l_0^2(w_i,\eta_a,T) + s_0^F(\tilde m_F^2, T, \mu),
  \end{aligned}
\end{equation}
where $\bar l_0^2$ is defined as in eq. \eqref{eq:threshold}, while $s_0^F$ reflects the different location of the Fermi surface for $\mu\neq 0$
\begin{align}
  &s_0^F(\tilde m_F^2, T, \mu) = \nonumber\\
  &= -\frac{t^2}{k}\int_{-\pi}^{\pi} \frac{d^2q}{(2\pi)^2} \tilde\partial_k
  \big\{\ln\cosh \frac{t\tilde y+\mu}{2T_k} + \ln\cosh \frac{t\tilde y-\mu}{2T_k} \big\},\\
  &\tilde y = \big(4(\cos q_1 + \cos q_2)^2+\tilde m_F^2\big)^{1/2}.\nonumber
\end{align}
The flow equations for the couplings $m_a^2$ and $\lambda_a$ in the bosonic potential can again be obtained by appropriate derivatives with respect to $\tilde\alpha$ (cf. eq. \eqref{eq:resc_ren_quant}).

The equation for the anomalous dimension becomes
%
%
\begin{eqnarray}
\eta_a &=& h_a^2 t k(\partial_kT_k)\Big[\frac{\partial}{\partial T_k}\frac{\partial}{\partial (l^2)}\\
&& \Big\{\frac{1}{T_k}\int_{-\pi}^{\pi} \frac{d^2q}{(2\pi)^2}
\frac{\tanh\tfrac{\epsilon(\Bs q)+\mu}{2T_k}
+\tanh\tfrac{\epsilon(\Bs q+l\hat e_1)-\mu}{2T_k}}
{(\epsilon(\Bs q)+\epsilon(\Bs q+l\hat e_1))/t}
\Big\}\Big]_{l=0}.\nonumber
\end{eqnarray}
For the Yukawa--coupling we obtain the same functional form of the beta functions as in eq. \eqref{eq:betafun_ha}. However, the fermionic propagator is now replaced by $p_F(Q) = (i\omega_F+\epsilon({\bf q})-\mu)/t$ (cf. eq. \eqref{eq:trunk_ferm}).

We have analysed the phase diagram of the Hubbard model for small values of the chemical potential $\mu$. The results are plotted in figure 13, where again we have chosen $U=3t$. The upper line shows the temperature $T_{pc}$ at which the bosonic mass vanishes in mean field theory (c.f. eq. \eqref{eq:bos_prop_korr} for $Q = \Pi$).
For the bosonisation we have chosen $U_\rho=0$.
For the small values of $\mu$ considered here the MFT approximation yields a second order phase transition such that the critical line indeed corresponds to a vanishing bosonic mass term. (In order to deal with first order phase transitions one would have to treat the bosonic potential in a more complicated truncation. We therefore restrict ourselves to small values of $\mu$ here.)
In the MFT-approximation $T_{pc}(\mu)$ therefore indicates the phase boundary and
$T_{pc}$ coincides with $T_c$.

\begin{figure}[htb!]
\centering
\psfrag{T}{$T_c/t$}
\psfrag{mu}{\hspace{-.5 em}$\mu/t$}
\includegraphics[scale=.65]{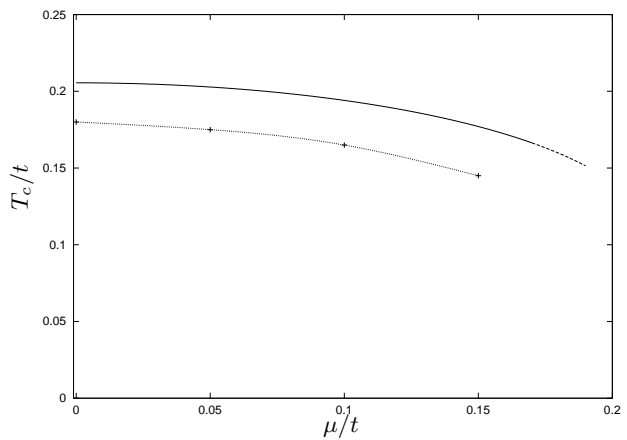}
\caption{Pseudocritical temperature $T_{pc}$ versus chemical potential $\mu$ for $U/t=3$ in the mean field approximation (above) and with flow equations (below).}
\label{fig:ayuk_Tc_mu}
\end{figure}

The lower curve in figure \ref{fig:ayuk_Tc_mu} shows the pseudocritical temperature $T_{pc}$ for various values of the chemical potential derived with the aid of the flow equations displayed above (again with $U=3t,U_\rho=0)$. One observes that the pseudocritical temperature is lowered as compared to the mean field result. This is precisely the effect of the bosonic fluctuations. Indeed, the bosonic fluctuations have the tendency to stabilize the symmetric phase -- they give a positive contribution to the mass term. This can explain why the pseudocritical temperature is lowered. Another effect is the influence
of the bosonic fluctuations on the running of the Yukawa coupling. The larger the Yukawa coupling the stronger is the destabilisation of the symmetric phase by the fermion fluctuations. This effect can lower or enhance $T_{pc}$, depending on the precise choice of the bosonisation. It will be discussed in more detail in sect. \ref{sec:remove_ambig}. The effect of the bosonic corrections is only moderate ($\sim 10\%$) for the parameters of our example. This can be very different for other choices of partial bosonisation, as visible in fig. \ref{fig:Tc_arho} for large $U_\rho/t$ where $T_{pc}$ is lowered by almost a factor of 2.


\section{Removing the mean field ambiguity}
\label{sec:remove_ambig}

In this chapter we want to investigate how well the inclusion of running couplings is able to solve the ambiguity with respect to the choice of Yukawa couplings in the bosonisation procedure, which was so annoying in the mean field calculation. For this purpose we add in our truncation the $\rho$--boson corresponding to fluctuations in the charge density. To keep things simple, however, we reduce the bosonic effective potential to a simple mass term for each boson and restrict our discussion to the symmetric phase.

For the full inverse propagator $P^a_{k}\big(Q)=P_a\big(Q)+R^a_{k}(Q)$ of the $\vec a$ boson we choose
%
%
\begin{equation}
 \begin{split}
  & P^a_{k}\big(Q=(\omega_B,\Bs q)\big) ={}\\
  &\qquad =\begin{cases}
    \bar m_{a\Lambda}^2 + R^a_{k} \equiv P^a_{\Lambda} & \text{for } \omega_B\neq 0\\
    Z_a [\Bs q]^2 + \bar m_{ak}^2 + R^a_{k} \equiv P^a_{k}(\Bs q) & \text{for } \omega_B = 0
  \end{cases}
 \end{split}
\end{equation}
and similarly for $P^\rho_{k}(Q)$ (however, we fix $Z_\rho=1$). The function $[\Bs q]^2$ is defined as below eq. \eqref{eq:trunk_bosprop}. This choice reflects the fact that in the one loop calculation we found that the $\omega_B=0$ mode is changed most. Furthermore, if we make an $\omega_B$--independent choice of the bosonic regulator, we are able to perform the Matsubara sums in the loops for the Yukawa--couplings, which drastically speeds up the numerics. We choose here a masslike cutoff $R^B_{k}=k^2$ for both $\vec a$ and $\rho$.
The fermionic kinetic part of the truncation is chosen as in section \ref{sec:trunc_antiferr}. Specifically, we restrict ourselves to nearest neighbour hopping. Furthermore we will only consider $\mu=0$. Also the Yukawa part is chosen as in section \ref{sec:trunc_antiferr}. Details of the flow equations for this truncation can be found in appendix \ref{ssec:af_and_cd_waves}.

This truncation is a very primitive one. It was chosen in order to highlight the dominant factor for the cure of the Fierz ambiguity, namely the inclusion of the bosonic fluctuations beyond mean field theory. The simple choice also makes the numerics relatively fast. We therefore do not expect very precise results but one should be able to see the general features of the flow.

The first check was to see if the one loop corrections to the fermionic coupling $U$ are reproduced by the above flow equations. For small $U$, and $T$ sufficiently above $T_c$, the perturbative expansion in $U$ is valid and one may compute the correction $\Delta U=cU^2$ in the purely fermionic language. Here
we compute the same quantity in the partially bosonized version. For this purpose we start at a large value of the cutoff $k$, follow the flow for a while and ``integrate out'' the bosons with their new couplings to obtain the new $U$.
As mentioned before, this amounts to a solution of the scalar field equations in presence of fermionic ``background fields''. As expected, it turns out that for large enough values of $k$ the flow indeed obeys the correct one loop equation. For small $U$ and large $T$ we recover the perturbative result. This holds independently of the choice of the bosonisation parameters $U_i$. This is of no surprise because of the one loop form of the flow equations and the fact that our truncation and rebosonisation systematically includes all terms contributing in order $U^2$. We emphasize, however, that the rebosonisation of sect. \ref{sec:rebos} is needed in order to obtain one loop accuracy \cite{Jae02}.

If we follow the flow towards smaller values of $k$, the purely fermionic coupling will not remain local but rather obtain a complicated momentum dependence. Furthermore, the loop calculations are no longer adequate as a comparison for the quality of the flow. We therefore need another quantity to investigate the invariance of the flow under different parameterisations of the bosonisation. For this purpose we have chosen the pseudocritical temperature $T_{pc}$ where the mass of the boson $\vec a$ vanishes.
\begin{figure}[!tb]
\centering
\psfrag{T}{$T_c/t$}
\psfrag{ar}{\hspace{-.5 em}$U_\rho/t$}
\includegraphics[scale=.65]{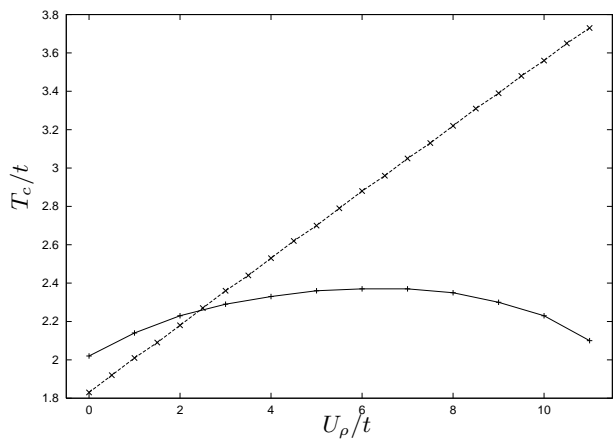}
\caption{Pseudocritical temperature $T_{pc}$ for different choices of the parameter $U_\rho$, calculated with flow equations (solid line) and in the mean field approximation (dotted line). The fermionic coupling is $U=12t=3U_m-U_\rho$. The Hartree--Fock (or Schwinger--Dyson) result
corresponds to the mean field value at $U_\rho=6t,$ i.e. $T_{pc}\approx 2.9t$.}
\label{fig:Tc_arho}
\end{figure}

In figure \ref{fig:Tc_arho} we plot the pseudocritical temperature $T_{pc}$ for the transition to the antiferromagnetic phase for different values of the parameters $U_m$ and $U_\rho$. Thereby we keep a fixed value of the four fermion coupling $U=12t=3U_m-U_\rho$. This graph has been calculated both with the flow equations and in the ``mean field approximation'', i.e. by searching for the zeroes of the bosonic mass in the fermionic one loop calculation of the two point function \footnote{The parameterisation independent Hartree--Fock results correspond to a critical temperature of approximately $T_c\approx 2.9t$ in this case.}. The fermionic cutoff was chosen to be $T_k^2=T^2+k^2$ for this plot. One observes that even for the solution of the flow equation the pseudocritical temperature still depends on the choice of the bosonisation. However, this dependence is relatively mild as compared to the mean field result and certainly due to our poor truncation. This shows that even a crude inclusion of the bosonic fluctuations leads to a substantial improvement over mean field theory as far as the Fierz ambiguity (choice of $U_i$) is concerned. The comparison in fig. \ref{fig:Tc_arho} with the Hartree--Fock result (corresponding to $U_\rho=U_m=6t$) also demonstrates the overall size of the change in $T_{pc}$ due to the bosonic fluctuations.

We mention that the scale independence of the unrenormalised Yukawa coupling $\bar{h}_a$ seen in figure 10 is also found in the present truncation for $U_\rho=0$. For other values of the parameters the unrenormalised Yukawa coupling indeed changes more substantially. In the present truncation the flow of $\bar m_a$ is influenced only through $\bar h_a$ (see \eqref{eq:flow_ma}). Therefore, if $\bar h_a$ is altered during the flow this will result in a change of the final value of $\bar m_a$ and hence the critical temperature. It is precisely the interplay of the running of $\bar{h}_a$ with
other effects that leads to the robustness of the flow equation result. In a sense, the flow
finds the ``correct value'' of $\bar{h}_a$ independently of the initial value for
$k\rightarrow\infty$. Indeed, the initial value depends
strongly on the choice of $U_m$ and this ``memory'' must be lost for physical quantities
to be independent of the choice of bosonisation.
We also observe that the ``Schwinger-Dyson value''
$U_\rho=U_m$ is not the choice which minimizes the effect of the bosonic fluctuations. This
rather happens for $U_\rho/t\approx 2.5$.

\end{fmffile}

\section{Conclusions}
\label{sec:conc}

The phase diagram of a high temperature superconductor shows many complicated features. At low doping and low temperature these materials are antiferromagnetic insulators. Increasing the concentration of electrons or holes turns them into a superconductor with exceptionally high transition temperatures as compared to ``conventional'' superconductors. Not much is known about the mechanism for the binding of electrons into Cooper pairs in these materials. Between doping concentrations leading to antiferromagnetic or superconducting behaviour one suspects a region in which many different degrees of freedom seem to play a role. The clarification of the basic degrees of freedom and their interplay in this pseudogap region still needs a lot of experimental and theoretical effort.

The common feature of all high temperature superconductors is their highly anisotropic structure composed of layers of copper oxide ($\rm CuO_2$) planes. The interesting properties of these materials and the mechanisms for generating them seem to be largely confined to these planes. The two dimensional Hubbard model is a simple attempt to capture this microscopic structure. The model assumes electrons that are able to tunnel from site to site on a lattice and feel a mutual screened Coulomb repulsion. Whether such an oversimplified model is able to reproduce the complex phase structure of a real high temperature superconductor or parts of it still has to be clarified. A lot of theoretical work has been dedicated to this task over the last years but so far the results are still inconclusive.

The aim of this paper is a computation of the phase diagram, the temperature dependence
of the antiferromagnetic order parameter and the critical behaviour of the correlation
length for the two dimensional Hubbard model with next neighbour coupling and small doping.
Although the quantitative results of this mainly theoretical investigation cannot be
directly compared with experimental results we believe that several universal features
actually hold independently of the details of the superconducting material. We attempt a
solution of the Hubbard model by means of renormalisation group (RG) equations. Earlier RG studies have already revealed the power of this technique in the context of the Hubbard model \cite{Zan97}-\cite{Gro01}. The use of a purely fermionic language, however, renders an investigation of the low temperature phase extremely difficult. We believe that it is favourable to include the interesting degrees of freedom more explicitly. This can be achieved by rewriting the original action of the Hubbard model in a partially bosonised form.

Already a simple mean field calculation in the partially bosonised Hubbard model leads to very encouraging results. We were able to reproduce \cite{Bai00} a phase diagram that closely resembles the one of a real high temperature superconductor. However, this simple approach also reveals an undesirable drawback of the bosonisation procedure. The couplings are not uniquely fixed by the reformulation procedure but there is an arbitrariness connected to different parameterisations of the coupling term. This ``Fierz ambiguity'' is also mirrored in the results. Even though the reformulation itself is exact, approximations break this parameterisation invariance and often lead to a strong dependence of the computed phase diagram on unphysical parameters.

In the mean field approximation the fluctuations of the bosonic fields are completely neglected. Taking the bosonic fields into account should dispose of or at least diminish the parameterisation dependence of the results. An inclusion of the effective bosonic fluctuations may be performed using renormalisation group equations. We base our investigation on truncations of an exact functional renormalisation group equation for the effective action. The partially bosonised theory then serves as a starting point for the flow of couplings. A one loop calculation serves as a guide for the formulation of suitable truncations schemes and also clarifies the relation between diagrams in the bosonised theory and the original fermionic formulation.

A first truncation deals with antiferromagnetic behaviour at and close to half filling. We are able to observe the breaking of the spin rotation symmetry and may follow the flow further into the ordered phase. We obtain a plausible explanation of why antiferromagnetic behaviour may be observed in the two dimensional model despite of the Mermin-Wagner theorem which forbids the breaking of a continuous symmetry in two dimensions for $T>0$. The antiferromagnetic order is lost only when averaging over extremely large spatial extensions, much beyond any practical macroscopic scale. (In a strict sense, observation of antiferromagnetism may be regarded as a finite size effect.)
We emphasize that effective long range order is a two-dimensional phenomenon which does not
need the existence of a third dimension in realistic materials \footnote{Of course, the third
dimension may nevertheless, play a role for some realistic cuprates.} (with weak coupling
between the plans).
For a low doping concentration we calculate a phase diagram that agrees well with other investigations.

A necessary check of the consistency of our truncation scheme concerns the question to which extend physical observables depend on the unphysical parameters introduced in the bosonisation procedure. We have studied the dependence of the pseudocritical temperature $T_{pc}$ on the unphysical ratio of Yukawa couplings $U_m/U_\rho$ (at fixed $U=3U_m-U_\rho$). Fig. 14 shows that the Fierz ambiguity of mean field theory is substantially reduced by the inclusion of the bosonic fluctuations, even in a very crude truncation. This result is encouraging and tests along this line should be performed for future more elaborate truncations.

Our main result is the temperature dependence of the antiferromagnetic order parameter $\kappa_a(T)$ shown in
fig. \ref{fig:ssb}. This determines the gap for the electrons according to
$\Delta_a=\sqrt{2h^2_a\kappa_a}t$. We also have given a detailed description of the
universal critical behaviour for temperatures near the critical temperature $T_c$, including
the temperature dependence of the correlation length for $T>T_c$ and an estimate of the range of
validity of the universal behaviour ($T\lesssim 0.8 T_{pc}$). In agreement with other methods
\cite{VD,Sca95,NMe} we find that $T_c$ is substantially lower than the Hartree-Fock result.

All our findings point to a simple Landau theory for the antiferromagnetic spin waves in the low
temperature phase
\begin{equation}\label{125}
S_L=\frac{1}{2}Z_a(T)\partial_i\vec{a}\partial_i\vec{a}
+\frac{1}{8}\bar{\lambda}_a(T)\left(\vec{a}^2-\frac{2T}{Z_a(T)t^2}\kappa_a(T)\right)^2.
\end{equation}
By variation with respect to $\vec{a}(x)$ one finds the field equations for static
spin wave configurations in thermal equilibrium
\begin{equation}\label{126}
\Delta\vec{a}=\frac{\bar{\lambda}_a}{2Z_a}\left(\vec{a}^2-
\frac{2T}{Z_at^2}\kappa_a\right)\vec{a}.
\end{equation}
The effective parameters $Z_a,\bar{\lambda}_a$ and $\kappa_a$ depend on $T$, the chemical
potential $\mu$ (doping) and the characteristic length scale of the configuration or
experimental probe $k^{-1}$. They can be directly extracted from the solution of the flow
equation. Since $Z_a$ and $\bar{\lambda}_a$ depend continuously on $\mu$ and are
positive for $\mu=0$ they will remain positive in a whole range of $\mu$. We
see no sign of instability towards a
spiral state for small nonzero doping \cite{SW1,SW2,MFH}.

In view of the self--consistency of our picture concerning
the Mermin--Wagner theorem and the Fierz ambiguity we believe that our exploration of
the low temperature phase is reliable. The quantitative accuracy can be improved
by future more extended truncations. This should confirm if all essential physical
ingredients are indeed taken into account in the present truncation. It will also
permit a more detailed study of the universal critical behaviour near the second order
phase transition and an extension of the phase diagram to larger doping.

Already at the present stage our investigation goes substantially beyond the results of a
Hartree-Fock or Schwinger-Dyson calculation. The additional effects of the bosonic
fluctuations that we have included in the present work lowers the critical temperature
by roughly factor two. This demonstrates the importance of these
fluctuations even for the rough features of the phase diagram. A robust picture of the
low temperature phase and the phase transition can only be reached if the fluctuation
effects of these collective bosonic degrees of freedom are taken into account properly.

Indeed, the functional renormalisation group in the partially bosonised formulation presented here has interesting prospects for understanding further features of the Hubbard model. It allows us to investigate directly the low temperature phases with nonvanishing order parameters. Further channels of possible spontaneous symmetry breaking can be described by additional bosonic degrees of freedom. This concerns, in particular, the boson responsible for $d$--wave superconductivity for large doping \cite{Bai00} which is briefly discussed in appendix D. Adding further bosonic channels also improves the resolution of the momentum dependence of the four fermion vertex. It remains to be seen how these methods can contribute to the fascinating theoretical challenge of establishing a quantitative phase diagram of the Hubbard model.

\appendix
\begin{widetext}
\section{four--fermion couplings}
\label{app:rebos}

In this appendix we compute the box diagrams shown in fig. 15.

\vspace{0.3cm}
\noindent
\vspace*{-0.3cm}
\begin{figure}[h]
\includegraphics[width=12cm]{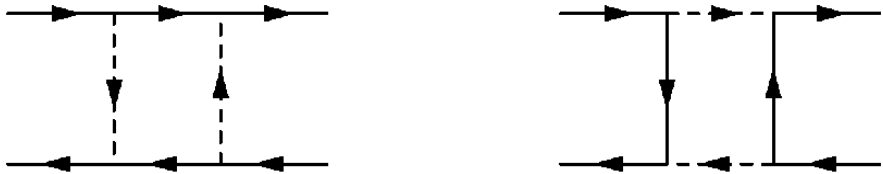}
\label{graph:fourferm_II}
\caption{Box diagrams generating four fermion interactions.}
\end{figure}
\vspace{0.3cm}
\noindent

\vspace{0.3cm}
\noindent
They generate four fermion interactions by the renormalisation flow, which will later be absorbed into effective bosonic interactions by rebosonisation (sect. \ref{sec:rebos}).

We choose the external momenta as
\begin{gather}
  A=Q, \quad B=Q-K, \quad C=Q', \quad D=Q'+K
\end{gather}
such that for the $\vec m$ boson
\begin{equation}
  \begin{split}
    \Delta\Gamma_\text{box} &= \sum_{KQQ'} \lambda_{QQ'}(K)\tilde {\vec m}_Q(K) \tilde {\vec m}_{Q'}(-K) + \cdots,\\
    \tilde {\vec m}_Q(K) &= \psi^*(Q) \vec\sigma \psi(Q+K).
  \end{split}
\end{equation}
For the $\rho$-boson the Pauli-matrices $\vec{\sigma}$ are omitted.
We evaluate $\lambda_{QQ'}(K)$ at
\begin{gather}
  Q=Q'=0, \quad \text{and} \quad
  \begin{cases}
    K=\Pi & \text{for the $\vec a$ boson}\\
    K=0   & \text{for the $\rho$ boson.}
  \end{cases}
\end{gather}
As an example, if only the $\vec a$ boson is present, we obtain
\begin{equation}
  \begin{split}
    \Delta\lambda_a = - \bar h_a^4 \sum_L P_a^{-1}(L) P_a^{-1}(\Pi-L) \;
    P_F^{-1}(-L)[P_F^{-1}(\Pi-L)-P_F^{-1}(L)].
  \end{split}
\end{equation}
For the evaluation of this expression leading to eqs. (\ref{eq:betafun_ha}),
(\ref{appB10}), (\ref{appB18}), (\ref{appB19}) one employs the identities
\begin{equation}
  \begin{split}
    (\sigma^i\sigma^j)_{\alpha\beta}(\sigma^i\sigma^j)_{\gamma\delta} &=
    3 \delta_{\alpha\beta}\delta_{\gamma\delta} - 2 \sigma^k_{\alpha\beta}\sigma^k_{\gamma\delta}, \\
    (\sigma^i\sigma^j)_{\alpha\beta}(\sigma^j\sigma^i)_{\gamma\delta} &=
    3 \delta_{\alpha\beta}\delta_{\gamma\delta} + 2 \sigma^k_{\alpha\beta}\sigma^k_{\gamma\delta}.
  \end{split}
\end{equation}

\section{Explicit form of the flow equations}

\subsection{Antiferromagnetism at half filling}

We display here the explicit flow equations for the mass term, order parameter and quartic scalar coupling for $\mu=0$. They are obtained by taking appropriate derivatives of eq. \eqref{eq:flow_u}. With the definitions $\eta_a = -\partial_t\ln Z_a$ (anomalous dimension) and $\overline{\sum_Q}f(Q) \equiv \sum_Q\Theta(k^2-{\hat Q}^2)f(Q)$ we get in the symmetric (SYM) regime
\begin{equation}
  \begin{split}
    k\partial_k m_a^2 =& -\tfrac 52 \frac{\lambda_a}{T}\overline{\sum_Q}\; \frac{2-\eta_a(1-{\hat Q}^2/k^2)}{(k/t)^2(1+m_a^2)^2} \\
    &+(\partial_k T_k) h_a^2 \frac{t^4}{2k T_k^3} \int_{-\pi}^{\pi} \frac{d^2q}{(2\pi)^2}
    \left\{\frac{\tanh y(0)}{y(0)}+\frac{1}{\cosh^2y(0)} \right\}\\
    &-(2-\eta_a)m_a^2,
  \end{split}
\end{equation}
\begin{equation}
  \begin{split}
    k\partial_k \lambda_a =& 11 \frac{\lambda_a^2}{T}\overline{\sum_Q}\; \frac{2-\eta_a(1-{\hat Q}^2/k^2)}{(k/t)^2(1+m_a^2)^3} \\
    &-(\partial_k T_k) h_a^4 \frac{t^6}{8k T_k^5} \int_{-\pi}^{\pi} \frac{d^2q}{(2\pi)^2}
    \left\{\frac{\tanh y(0)}{y^3(0)}+\frac{2y(0)\tanh y(0)-1}{y^2(0)\cosh^2y(0)} \right\}\\
    &-2(1-\eta_a)\lambda_a.
  \end{split}
\end{equation}
In the broken (SSB) regime we find
\begin{equation}
  \begin{split}
    k\partial_k \kappa_a =& \frac{1}{2T}\overline{\sum_Q}\; \frac{2-\eta_a(1-{\hat Q}^2/k^2)}{(k/t)^2}
    \left(\frac{3}{(1+2\lambda_a\kappa_a)^2}+2\right) \\
    &-(\partial_k T_k) \frac{h_a^2}{\lambda_a} \frac{t^4}{2k T_k^3} \int_{-\pi}^{\pi} \frac{d^2q}{(2\pi)^2}
    \left\{\frac{\tanh y(\kappa_a)}{y(\kappa_a)}+\frac{1}{\cosh^2y(\kappa_a)} \right\}\\
    &-\eta_a \kappa_a,
  \end{split}
\end{equation}
\begin{equation}
  \label{eq:flow_lambda_ssb}
  \begin{split}
    k\partial_k \lambda_a =& \frac{\lambda_a^2}{T}\overline{\sum_Q}\; \frac{2-\eta_a(1-{\hat Q}^2/k^2)}{(k/t)^2}
    \left(\frac{9}{(1+2\lambda_a\kappa_a)^3}+2\right)\\
    &-(\partial_k T_k) h_a^4 \frac{t^6}{8k T_k^5} \int_{-\pi}^{\pi} \frac{d^2q}{(2\pi)^2}
    \left\{\frac{\tanh y(\kappa_a)}{y^3(\kappa_a)}+\frac{2y(\kappa_a)\tanh y(\kappa_a)-1}{y^2(\kappa_a)\cosh^2y(\kappa_a)} \right\}\\
    &-2(1-\eta_a)\lambda_a,
  \end{split}
\end{equation}
where
\begin{equation}
  y(\kappa_a) = \frac{1}{2T_k}\sqrt{\epsilon^2(\Bs q)+2t^2h_a^2\kappa_a}.
\end{equation}
If $k$ obeys both $k<2\pi T$ and $k < \pi$ we are able to evaluate the $\overline\sum_Q$--sum and find $\overline\sum_Q (2-\eta_a(1-{\hat Q}^2/k^2))=k^2T(4-\eta_a)/(8\pi)$.

\subsection{Antiferromagnetism for nonzero doping}

For $\mu\neq 0$ the flow equations for the couplings $m_a^2$ and $\lambda_a$ in the bosonic potential can be obtained by appropriate derivatives of \eqref{eq:flow_u_mu} with respect to $\tilde\alpha$
\begin{equation}
  \begin{split}
    k\partial_k m_a^2 =& -\tfrac 52 \frac{\lambda_a}{T} \overline{\sum_Q}\; \frac{2-\eta_a(1-{\hat Q}^2/k^2)}{(k/T)^2(1+m_a^2)^2} \\
    &+(\partial_k T_k) h_a^2 \frac{t^4}{2k T_k^3} \int_{-\pi}^{\pi} \frac{d^2q}{(2\pi)^2}
    \left\{\frac{\tanh \tilde y(\mu)}{\tilde y(0)}
      +\frac{\tilde y(\mu)/\tilde y(0)}{\cosh^2 \tilde y(\mu)} \right\}\\
    &-(2-\eta_a)m_a^2,
  \end{split}
\end{equation}
\begin{equation}
  \begin{split}
    k\partial_k \lambda_a =& 11 \frac{\lambda_a^2}{T} \overline{\sum_Q}\; \frac{2-\eta_a(1-{\hat Q}^2/k^2)}{(k/T)^2(1+m_a^2)^3} \\
    &-(\partial_k T_k) h_a^4 \frac{t^6}{8k T_k^5} \int_{-\pi}^{\pi} \frac{d^2q}{(2\pi)^2}
    \left\{\frac{\tanh \tilde y(\mu)}{\tilde y^3(0)}
      +\frac{2\tilde y(\mu)\tanh \tilde y(\mu)-\tilde y(-\mu)/\tilde y(0)}{\tilde y^2(0)\cosh^2 \tilde y(\mu)}
    \right\}\\
    &-2(1-\eta_a)\lambda_a,
  \end{split}
\end{equation}
with
\begin{equation}
  \tilde y(\mu) = (\epsilon(\Bs q)-\mu)/(2T_k).
\end{equation}

For the flow equation of the Yukawa coupling we evaluate eq. \eqref{eq:betafun_ha} with $p_F(Q) = (i\omega_F+\epsilon({\bf q})-\mu)/t$. One finds
\begin{equation}
  k\partial_k h_a^2 = h_a^2 \eta_a + \beta_{h^2_a}^{(d)} + \beta_{h^2_a}^{(rb)},
\end{equation}
\begin{equation}\label{appB10}
  \begin{split}
    \beta_{h^2_a}^{(d)}
    =& - 2 h_a^4 \frac{t^4}{k^2 T} k\tilde\partial_k\sum_Q\big\{ P_{F1}^{-1}(Q)p_a(Q) \big\}\\
    =& - 2 h_a^4 \frac{t^4}{k^2 T} \sum_Q\bigg\{-(k\partial_k P_{F1}^{-1}(Q))\frac{1}{Q_k^2/k^2+m_a^2}
    + P_{F1}^{-1}(Q) \frac{(2-\eta_a(1-{\hat Q}^2/k^2))\Theta(k^2-{\hat Q}^2)}{(1+m_a^2)^2}
    \bigg\},
    \\
    \beta_{h^2_a}^{(rb)}
    =& - 4m_a^2 h_a^4 \frac{t^4}{k^2 T} k\tilde \partial_k\sum_Q\big\{P_{F2}^{-1}(Q) p_a^{-1}(Q)p_a^{-1}(Q+\Pi) \big\} \\
    =& - 4m_a^2 h_a^4 \frac{t^4}{k^2 T} \sum_Q\bigg\{(k\partial_k P_{F2}^{-1}(Q))\frac{1}{Q_k^2/k^2+m_a^2}
    - 2 P_{F2}^{-1}(Q)\frac{(2-\eta_a(1-{\hat Q}^2/k^2))\Theta(k^2-{\hat Q}^2)}{(1+m_a^2)^2}
    \bigg\}\\
    &\hspace{5.cm} \times \frac{1}{(Q+\Pi)_k^2/k^2+m_a^2}
  \end{split}
\end{equation}
with
\begin{eqnarray}
  P_{F1}^{-1}(Q) &=& \frac{\omega_F^2+\epsilon(\Bs q)^2-\mu^2}
  {(\omega_F^2+\epsilon(\Bs q)^2-\mu^2)^2+4\omega_F^2\mu^2},\nonumber\\
  P_{F2}^{-1}(Q) &=& \frac{\omega_F^2+\epsilon(\Bs q)^2}
  {(\omega_F^2+(\epsilon(\Bs q)+\mu)^2)^2 (\omega_F^2+(\epsilon(\Bs q)-\mu)^2)^2}.
\end{eqnarray}

\subsection{Antiferromagnetism and charge density waves}
\label{ssec:af_and_cd_waves}

The running of the {\bf anomalous dimension} $\eta_a$ can be inferred directly from \eqref{eq:eta_sym}.
The flow of the {\bf masses} is governed by a fermionic loop (the bosonic loop does not contribute as we have neglected the quartic bosonic coupling) and reads
\begin{equation}
  \label{eq:flow_ma}
  \begin{split}
    \partial_k \bar m_a^2 = (\partial_k T_k) \bar h_a^2
    \frac{T}{2 T_k^3} \int_{-\pi}^{\pi} \frac{d^2q}{(2\pi)^2}
    \left\{\frac{\tanh y}{y}+\frac{1}{\cosh^2y} \right\},
  \end{split}
\end{equation}
\begin{equation}
  \begin{split}
    \partial_k \bar m_\rho^2 = (\partial_k T_k) \bar h_\rho^2
    \frac{T}{T_k^3} \int_{-\pi}^{\pi} \frac{d^2q}{(2\pi)^2}
    \left\{\frac{1-y \tanh y}{\cosh^2y} \right\},
  \end{split}
\end{equation}
where $y=\frac{\epsilon_{\Bs q}}{2T_k}$.

The running of the {\bf Yukawa couplings} is governed by
\begin{equation}
  k\partial_k \bar h_i = \beta_{\bar h_i}^{(d)} + \beta_{\bar h_i}^{(rb)}
\end{equation}
with the beta--functions
\begin{equation}
  \begin{split}
    \beta_{\bar h_a}^{(d)} = -\bar h_a k\tilde\partial_k \sum_L P_F^{-1}(L+\Pi) P_F^{-1}(L)
    \big\{ \bar h_a^2 P_a^{-1}(L) - \bar h_\rho^2 P_\rho^{-1}(L) \big\},
  \end{split}
\end{equation}
\begin{equation}
  \begin{split}
    \beta_{\bar h_\rho}^{(d)} = \bar h_\rho k\tilde\partial_k \sum_L P_F^{-2}(L)
    \big\{ 3 \bar h_a^2 P_a^{-1}(L) + \bar h_\rho^2 P_\rho^{-1}(L) \big\}
  \end{split}
\end{equation}
and
\begin{equation}\label{appB18}
  \begin{split}
    \beta_{\bar h_a}^{(rb)} =& \frac{\bar m_a^2}{2\bar h_a} k\tilde\partial_k \sum_L \Big\{
    2 \bar h_a^4 P_a^{-1}(L) P_a^{-1}(\Pi-L) \; P_F^{-1}(-L)[P_F^{-1}(\Pi-L)-P_F^{-1}(L)] \\
    & \hspace{1.5cm} + \bar h_a^2 \bar h_\rho^2\big( P_a^{-1}(L)P_\rho^{-1}(\Pi-L)+P_\rho^{-1}(L)P_a^{-1}(\Pi-L)\big)
    \times P_F^{-1}(-L)[P_F^{-1}(\Pi-L)+P_F^{-1}(L)]
    \Big\},
  \end{split}
\end{equation}
\begin{equation}\label{appB19}
  \begin{split}
    \beta_{\bar h_\rho}^{(rb)} =& \frac{\bar m_\rho^2}{2\bar h_\rho} k\tilde\partial_k \sum_L \big\{
    \big(3 \bar h_a^4 P_a^{-2}(L) + \bar h_\rho^4 P_\rho^{-2}(L) \big) \; P_F^{-1}(-L)[P_F^{-1}(-L)+P_F^{-1}(L)]
    \big\}.
  \end{split}
\end{equation}
In a more explicit form they read
\begin{equation}
  \begin{split}
    \partial_k \bar h_a =& \bar h_a T \int_{-\pi}^{\pi} \frac{d^2q}{(2\pi)^2}
    \tilde\partial_k \bigg[
    \bigg(
    \frac{1}{(\pi T_k)^2+\epsilon_{\Bs q}^2}
    \bigg\{
    \Big(\frac{\bar h_{ak}^2}{P_{ak}(\Bs q)} - \frac{\bar h_{a\Lambda}^2}{P_{a\Lambda}}\Big)
    - \Big(\frac{\bar h_{\rho k}^2}{P_{\rho k}(\Bs q)} - \frac{\bar h_{\rho\Lambda}^2}{P_{\rho\Lambda}}\Big) \bigg\}
    +\frac{\tanh y}{4T_k^2 y}\bigg\{
    \frac{\bar h_{a\Lambda}^2}{P_{a\Lambda}} - \frac{\bar h_{\rho\Lambda}^2}{P_{\rho\Lambda}} \bigg\}
    \bigg)\bigg]\\
    &-\frac{2T\bar m_{ak}^2}{\bar h_a} \int_{-\pi}^{\pi} \frac{d^2q}{(2\pi)^2}
    \tilde\partial_k \bigg[
    \bigg(\frac{1}{(\pi T_k)^2+\epsilon_{\Bs q}^2}
    \bigg\{
    \frac{\bar h_{ak}^4}{P_{ak}(\Bs q) P_{ak}(\Bs\pi-\Bs q)}-\frac{\bar h_{a\Lambda}^4}{P_{a\Lambda} P_{a\Lambda}}
    \bigg\}
    +\frac{\tanh y}{4T_k^2 y}
    \frac{\bar h_{a\Lambda}^4}{P_{a\Lambda}^2}
    \bigg)
    \bigg]
  \end{split}
\end{equation}
with $y=\frac{\epsilon_{\Bs q}}{2T_k}$. Similarly we find for $\bar h_\rho$
\begin{equation}
  \begin{split}
    \partial_k \bar h_\rho =& \bar h_\rho T \int_{-\pi}^{\pi} \frac{d^2q}{(2\pi)^2}
    \tilde\partial_k \bigg[
    \bigg(
    \frac{\epsilon_{\Bs q}^2-(\pi T_k)^2}{[\epsilon_{\Bs q}^2+(\pi T_k)^2]^2}\bigg\{
    \Big(\frac{\bar h_{\rho k}^2}{P_{\rho k}(\Bs q)} - \frac{\bar h_{\rho\Lambda}^2}{P_{\rho\Lambda}}\Big)
    +3\Big(\frac{\bar h_{ak}^2}{P_{ak}(\Bs q)} - \frac{\bar h_{a\Lambda}^2}{P_{a\Lambda}}\Big)
    \bigg\} 
    -\frac{1}{4T_k^2 \cosh^2y}\bigg\{
    \frac{\bar h_{\rho\Lambda}^2}{P_{\rho\Lambda}} + 3 \frac{\bar h_{a\Lambda}^2}{P_{a\Lambda}} \bigg\}
    \bigg)\bigg]\\
    &+\frac{T\bar m_{\rho k}^2}{\bar h_\rho}\int_{-\pi}^{\pi} \frac{d^2q}{(2\pi)^2}
    \tilde\partial_k \bigg[
    \bigg(\frac{\epsilon_{\Bs q}^2}{((\pi T_k)^2+\epsilon_{\Bs q}^2)^2}
    \bigg\{
    3\big(\frac{\bar h_{ak}^4}{P_{ak}^2(\Bs q)}-\frac{\bar h_{a\Lambda}^4}{P_{a\Lambda}^2}\big)
    +\big(\frac{\bar h_{\rho k}^4}{P_{\rho k}^2(\Bs q)}-\frac{\bar h_{\rho\Lambda}^4}{P_{\rho\Lambda}^2}\big)
    \bigg\}\\
    &\hspace{7cm}+\frac{1}{8T_k^2}\big(\frac{\tanh y}{y}-\frac{1}{\cosh^2 y}\big)
    \bigg\{
    3\frac{\bar h_{ak}^4}{P_{a\Lambda}^2}+\frac{\bar h_{\rho\Lambda}^4}{P_{\rho\Lambda}^2}
    \bigg\}\bigg)
    \bigg].
  \end{split}
\end{equation}

\end{widetext}
\section{Fermionic regulator}

\begin{figure*}[tb!]
\centering
\psfrag{F}{$F$}
\psfrag{x}{\raisebox{-2mm}{$q_1$}}
\psfrag{y}{\hspace{-3mm}\raisebox{-2mm}{$q_2$}}
\psfrag{a}{\raisebox{-2mm}{$\scriptstyle \frac \pi 2$}}
\psfrag{b}{\raisebox{-2mm}{$\scriptstyle \pi$}}
\psfrag{-a}{\raisebox{-2mm}{$\scriptstyle -\frac \pi 2$}}
\psfrag{-b}{\raisebox{-2mm}{$\scriptstyle -\pi$}}
\psfrag{n}{\raisebox{-2mm}{$\scriptstyle 0$}}
\includegraphics[scale=0.65]{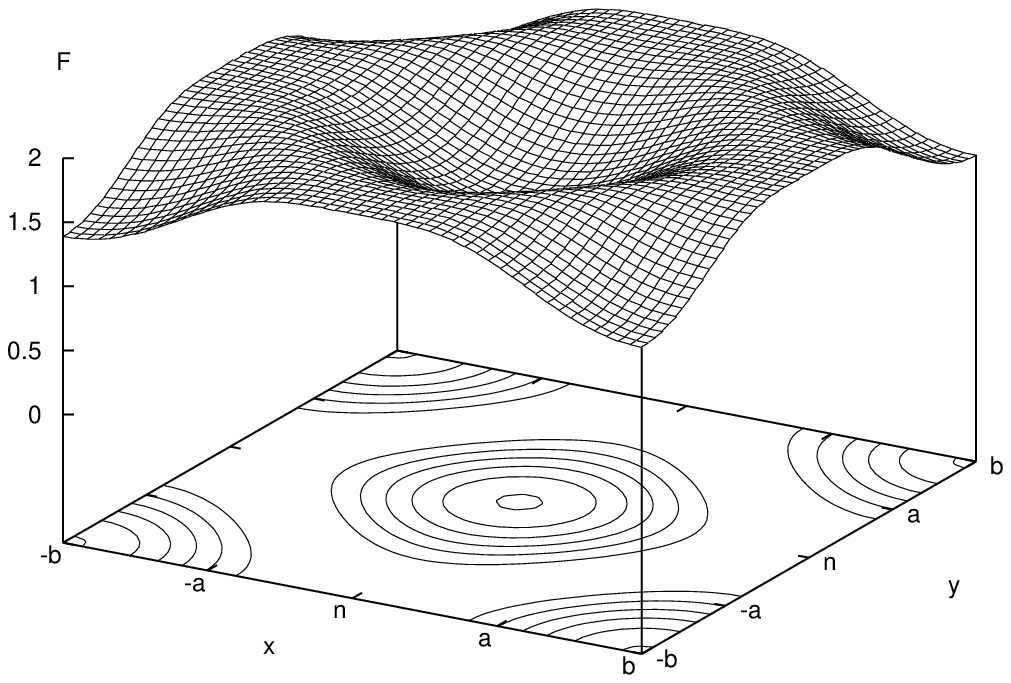}
\hspace{-1.8cm}
\includegraphics[scale=0.65]{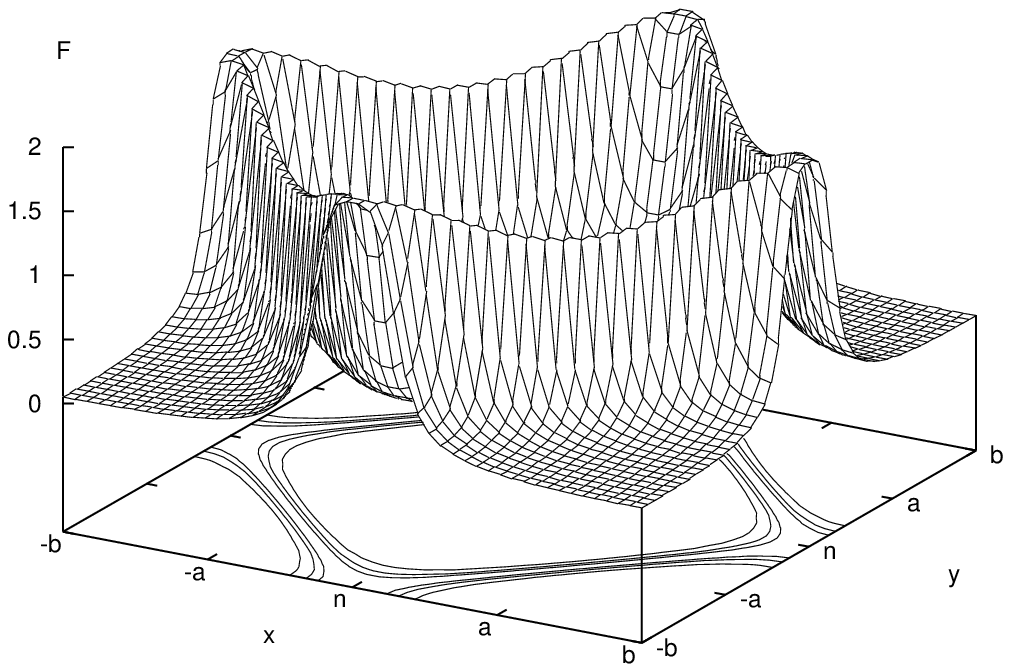}
\caption{Fermionic regulator. We plot the function $F(q_1,q_2)$ defined in \eqref{eq:integrand_m} for $T_k/t=5$ (left) and $T_k/t=1/5$ (right).}
\label{fig:integrand_m}
\end{figure*}

Let us take a look at how the fermionic regulator function works. For this purpose we investigate the integrand of the fermionic part of the flow equation for the bosonic mass (\ref{appB10}), i.e. the function
\begin{equation}
  \label{eq:integrand_m}
  F(q_1,q_2) = \frac{\tanh y}{y} + \frac{1}{\cosh^2 y},\qquad y = \frac{2t(\cos q_1 + \cos q_2)}{T_k}.
\end{equation}
If the flow is properly implemented $F(q_1,q_2)$ should be dominated by momenta close to the
Fermi surface for small $T_k/t$, since we only change the size of the shell of
fluctuations around the Fermi-surface as $k$ is lowered. In fig. 16 we plot $F(q_1,q_2)$
for different values of $T_k/t$. The left part of fig. 16 shows $F(q_1,q_2)$ for fairly large values of $T_k$ while the right part is for low $T_k$. Observe that the contribution to the integral comes from narrower regions around the Fermi--surface the smaller $T_k$ becomes. This was exactly what was intended by the regulator.

\section{Cooper pairs in the $d$-wave channel}

The Hubbard interaction may be decomposed into fermion bilinears in many different ways.
To display a few, we define fermion bilinears corresponding to charge density,
magnetisation and Cooper pairs in different channels
\begin{eqnarray}
  \tilde\rho (X) &\equiv& \tilde\rho_i = \hat\psi^*_i\hat\psi_i, \\
  \tilde {\vec m} (X) &\equiv& \tilde{\vec m}_i = \hat\psi^*_i \vec\sigma \hat\psi_i, \nonumber\\
  \tilde s(X) &\equiv& \tilde s_i = \hat\psi_i\epsilon\hat\psi_i,\quad \tilde s^*(X)
  \equiv \tilde s^*_i = -\hat\psi^*_i\epsilon\hat\psi^*_i, \nonumber\\
  \tilde c_x(X) &\equiv& \tilde c_{xi} = \hat\psi_i\epsilon\hat\psi_{i+\hat e_x},
  \quad \tilde c_x^*(X) \equiv \tilde c_{xi}^* = -\hat\psi^*_{i+\hat e_x}
  \epsilon\hat\psi^*_i ,\nonumber
\end{eqnarray}
where $\epsilon$ is the two dimensional completely antisymmetric tensor ($\epsilon=i\sigma^2$) and $\hat e_x$ is the unit vector in $x$-direction. We also define $\tilde c_y$ similar to $\tilde c_x$. With these definitions we may rewrite the interaction term as follows
\begin{equation}
  (\hat\psi^*_i\hat\psi_i)^2 = \tilde\rho_i^2 = -\tfrac 13 \tilde{\vec m}_i^2 = -{\tilde{\vec m}}_{3,i}^2 = \tfrac 12 \tilde s^*_i\tilde s_i
\end{equation}
and further note the identity
\begin{equation}
  -\tilde\rho_i \tilde\rho_{i+\hat e_x} + \tilde{\vec m}_i \tilde{\vec m}_{i+\hat e_x} + 2 \tilde{c_x^*}_i\tilde {c_x}_i = 0
\end{equation}
and similar for $x\to y$.

Let us now introduce auxiliary fields $\hat B = (\hat\rho,\hat{\vec m},\hat s, \hat c,\hat c_y)$ and add a term quadratic in these fields to the action such that the four fermion interaction is just cancelled:
\begin{eqnarray}\label{eq:S_bosonised}
&&S[\hat\psi,\hat\psi^*,\hat B] = S[\hat\psi,\hat\psi^*]
+\Delta S_{int}[\hat\psi,\hat\psi^*,\hat B],\\
\label{eq:deltaS}
&&\Delta S_{int} = \sum_X\bigg\{ \tfrac 12 U_\rho (\hat\rho_i - \tilde\rho_i)^2
+ \tfrac 12 U_m (\hat{\vec m}_i - \tilde{\vec m}_i)^2\nonumber\\
&&+ U_s (\hat s_i^*-\tilde s_i^*)(\hat s_i-\tilde s_i)
+ U_x \big[(\hat c_{xi}^*-\tilde c_{xi}^*)(\hat c_{xi}-\tilde c_{xi}) \nonumber\\
&&- \tfrac 12(\hat\rho_i-\tilde\rho_i)(\hat\rho_{i+\hat e_x}-\tilde\rho_{i+\hat e_x})\\
&&+ \tfrac 12(\hat{\vec m}_i-\tilde{\vec m}_i)(\hat{\vec m}_{i+\hat e_x}
-\tilde{\vec m}_{i+\hat e_x})\big]+ U_y [x \to y]
\bigg\}. \nonumber
\end{eqnarray}
Restricting the couplings to the range
\begin{gather}
  \label{eq:couplings}
  \begin{split}
    & U_i > 0, \\
    & U_\rho, U_m > U_x+U_y,
  \end{split}
\end{gather}
ensures that the auxiliary fields are Gaussian and can be integrated out after a shift of variables. Imposing
\begin{equation}\label{D6AA}
3U_m - U_\rho - 2U_s = U
\end{equation}
furthermore ensures that the resulting four fermion interaction in $\Delta S_{int}$ cancels precisely the one in the original action. These conditions thus ensure that the partially bosonised partition function
\begin{equation}\label{D7A}
  Z[\eta,\eta^*] = \int {\cal D} (\hat\psi^*,\hat\psi,\hat B)
  \exp \Big( - S[\hat\psi,\hat\psi^*,\hat B] + \eta^* \hat\psi + \eta \hat\psi^* \Big)
\end{equation}
is indeed equivalent to \eqref{eq:Zorig}.

Collecting the terms in \eqref{eq:S_bosonised} we see that we are now dealing with a theory of fermions coupled to bosons via a Yukawa interaction. In Fourier space the partially bosonised action reads
\begin{eqnarray}\label{eq:S_bos_four}
&&S[\hat\psi,\hat\psi^*,\hat B] = S_{kin}[\hat\psi,\hat\psi^*,\hat B]
+ S_Y[\hat\psi,\hat\psi^*,\hat B], \\
&&S_{kin} = \sum_Q\Big\{ \hat\psi^*(Q) [ i\omega_Q - \mu+\epsilon(q)]\hat\psi(Q)\nonumber\\
&&+ \tfrac{1}{2}(U_\rho-U_x\cos q_x-U_y\cos q_y)\hat\rho(-Q)\hat\rho(Q)\nonumber \\
&&+ \tfrac{1}{2}(U_m+U_x\cos q_x+U_y\cos q_y)\hat{\vec m}(-Q)\hat{\vec m}(Q)\nonumber \\
&&+ U_s \hat s^*(Q)\hat s(Q)
+ U_x \hat c_x^*(Q)\hat c_x(Q) + U_y \hat c_y^*(Q)\hat c_y(Q)\Big\}\nonumber
\end{eqnarray}
where for next neighbor $(t)$ and diagonal neighbor $(t^\prime)$ couplings one has
\begin{equation}\label{D9}
\epsilon(q)= - 2t(\cos q_x+\cos q_y) - 4t'\cos q_x\cos q_y].
\end{equation}
The Yukawa interaction reads
\begin{eqnarray}\label{D8AA}
&&S_Y = -\sum_{KQQ'}\bigg[\delta(K-Q+Q')\nonumber\\
&&\bigg\{ (U_\rho-U_x\cos k_x-U_y\cos k_y) \hat\rho(K)
\hat\psi^*(Q)\hat\psi(Q') \nonumber\\
&&+ (U_m+U_x\cos k_x+U_y\cos k_y)\hat{\vec m}(K)
\hat\psi^*(Q)\vec\sigma\hat\psi(Q')
\bigg\}\nonumber\\
&&+\delta(K-Q-Q')\nonumber\\
&&\bigg\{
U_s[\hat s^*(K)\hat\psi(Q)\epsilon\hat\psi(Q')-\hat s(K)\hat\psi^*(Q)
\epsilon\hat\psi^*(Q')]\nonumber\\
&&
+U_x\cos\tfrac{q_x-q'_x}{2}[\hat c_x^*(K)\hat\psi(Q)\epsilon\hat\psi(Q')\nonumber\\
&&-\hat c_x(K)\hat\psi^*(Q)\epsilon\hat\psi^*(Q')]\nonumber\\
&&+U_y\cos\tfrac{q_y-q'_y}{2}[\hat c_y^*(K)\hat\psi(Q)\epsilon\hat\psi(Q')\nonumber\\
&&-\hat c_y(K)\hat\psi^*(Q)\epsilon\hat\psi^*(Q')]\bigg\}\bigg].
\end{eqnarray}
Here we have used the Fourier transforms \eqref{eq:four} for the fermions and
\begin{equation}
  \label{eq:fourtrans_s}
  \hat\chi(X) = \sum_Q e^{iQX}\hat\chi(Q), \quad \hat\chi^*(X) = \sum_Q e^{-iQX}\hat\chi^*(Q)
\end{equation}
for $\hat\chi^{(*)}=(\hat \rho, \hat m, \hat s^{(*)})$, while for $\hat c,\hat c^*$ we use:
\begin{equation}
  \label{eq:fourtrans_c}
  \begin{split}
    \hat c_x(X) &= \sum_Q e^{i(QX+q_x/2)}\hat c_x(Q), \\
    \hat c^*_x(X) &= \sum_Q e^{-i(QX+q_x/2)}\hat c^*_x(Q)
  \end{split}
\end{equation}
and similar for $c^{(*)}_y$.

In order to get an operator that has $d$--wave symmetry we perform the transformation of variables
\begin{equation}
  {\hat e = (\hat c_x+\hat c_y), \atop \hat d = (\hat c_x-\hat c_y),}  \qquad
  {\hat c_x = \tfrac 12(\hat e+\hat d), \atop \hat c_y = \tfrac 12(\hat e-\hat d),}
\end{equation}
and similar for $\hat e^*, \hat d^*$. If we insert this variable transformation into the action \eqref{eq:S_bos_four} and put $U_x=U_y=U_c$ we obtain for the $e$ and $d$ dependent part of the action:
\begin{eqnarray}\label{eq:S_ed}
&&S^{e,d} = \sum_Q\tfrac 12U_c\{\hat e^*(Q)\hat e(Q)+ \hat d^*(Q)\hat d(Q)\}\\
&&-\sum_{KQQ'}\delta(K-Q-Q')
\Big\{\tfrac{U_c}{2} (\cos\tfrac{q_x-q'_x}{2}+\cos\tfrac{q_x-q'_x}{2})\nonumber\\
&&\times[\hat e^*(K)\hat\psi(Q)\epsilon\hat\psi(Q')-\hat e(K)\hat\psi^*(Q)\epsilon\hat\psi^*(Q')]\nonumber\\
&&+\tfrac{U_c}{2} (\cos\tfrac{q_x-q'_x}{2}-\cos\tfrac{q_x-q'_x}{2})\nonumber\\
&&\times[\hat d^*(K)\hat\psi(Q)\epsilon\hat\psi(Q')-\hat d(K)\hat\psi^*(Q)\epsilon\hat\psi^*(Q')]
\Big\}.\nonumber
\end{eqnarray}
Integrating out the bosons is equivalent to inserting the solutions of their field equations $\delta S[\hat\psi,\hat\psi^{*},\hat B]/\delta \hat B=0$ (i.e. the saddle point) into the action $S[\hat\psi,\hat\psi^{*},\hat B]$. For the boson $\hat d$ the solution is
\begin{equation}
  \label{eq:def_tilde_d}
  \hat d(Q) = \tilde c_x(Q) - \tilde c_y(Q)=\tilde{d}(Q),
\end{equation}
as expected from the construction of $\hat d$. In the fermionic theory
it is thus this combination that is represented by the boson $\hat d$.

\begin{figure}[!tb]
  \centering
  \unitlength1mm
  \includegraphics[scale=1.]{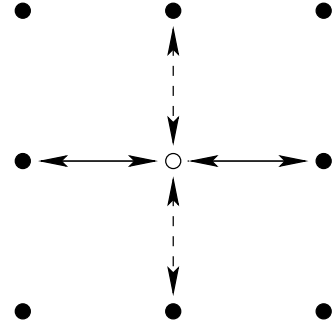}
  \caption{The $\Bs q = \Bs 0$ mode of $\tilde d(Q)$ in its local form. Solid and dashed lines indicate that two fermion operators on neighbouring lattice sites are connected with positive or negative sign respectively.}
  \label{fig:d-wave}
\end{figure}
Let us take a look at the $\Bs q = (0,0)$ mode of $\tilde d(Q)$, i.e. at a spatially homogeneous field. From \eqref{eq:def_tilde_d} we know that it is a superposition of stripes along the $x$-- and $y$--axis added with opposite signs. To find a ``local'' expression we rewrite
\begin{eqnarray}
\tilde d(Q&=&0) = \sum_K (\cos k_x - \cos k_y) \hat\psi(K)\epsilon\hat\psi(-K) \\
&=&\tfrac 12 \sum_X\big\{
\hat\psi(X)\epsilon\hat\psi(X+\hat e_x) + \hat\psi(X)\epsilon\hat\psi(X-\hat e_x)\nonumber\\
&&\hspace{1.5cm}-\hat\psi(X)\epsilon\hat\psi(X+\hat e_y) - \hat\psi(X)\epsilon\hat\psi(X-\hat e_y) \big\}\nonumber
\end{eqnarray}
so that at each lattice site we find an operator of the form shown in figure 17.
We see that indeed this boson may serve as a lattice representation of a $d_{x^2-y^2}$
wave as it changes its sign under rotation by $90^\circ$ but not under reflection
at the $x$ or $y$ axes (see also \cite{Sca95} for a more extensive classification). The
mean field approach including the boson $\hat{d}$ is discussed in
\cite{Bai00}.




\end{document}